\documentclass[manuscript]{aastex63}
\usepackage{ragged2e}
\usepackage{mathtools}

\providecommand{\e}[1]{\ensuremath{\times 10^{#1}}}

\submitjournal{apj}

\shortauthors{Salman et al.}

\begin{document}

\title{Categorization of Coronal Mass Ejection Driven Sheath Regions: \textbf{Characteristics} of STEREO Events}

\correspondingauthor{Tarik M. Salman}
\email{ts1090@wildcats.unh.edu}

\author[0000-0001-6813-5671]{Tarik M. Salman}
\affiliation{Space Science Center and Department of Physics \\
University of New Hampshire \\
Durham, NH 03824, USA}

\author[0000-0002-1890-6156]{No{\'e} Lugaz}
\affiliation{Space Science Center and Department of Physics \\
University of New Hampshire \\
Durham, NH 03824, USA}

\author[0000-0002-9276-9487]{Reka M. Winslow}
\affiliation{Space Science Center and Department of Physics \\
University of New Hampshire \\
Durham, NH 03824, USA}

\author[0000-0001-8780-0673]{Charles J. Farrugia}
\affiliation{Space Science Center and Department of Physics \\
University of New Hampshire \\
Durham, NH 03824, USA}

\author[0000-0002-6849-5527]{Lan K. Jian}
\affiliation{Heliophysics Science Division \\
NASA Goddard Space Flight Center \\
Greenbelt, MD 20771, USA}

\author[0000-0003-3752-5700]{Antoinette B. Galvin}
\affiliation{Space Science Center and Department of Physics \\
University of New Hampshire \\
Durham, NH 03824, USA}

\begin{abstract}

\justify

We present a comprehensive statistical analysis of 106 sheath regions driven by coronal mass ejections (CMEs) and measured near 1 AU. Using data from the STEREO probes, this extended analysis focuses on two discrete categorizations. In the first categorization, we investigate how the generic features of sheaths change with their potential formation mechanisms (propagation and expansion sheaths), namely, their associations with magnetic ejectas (MEs) which are primarily expanding or propagating in the solar wind. We find propagation sheaths to be denser and driven by stronger MEs, whereas expansion sheaths are faster. Exploring the temporal profiles of these sheaths with a superposed epoch technique, we observe that most of the magnetic field and plasma signatures are more elevated in propagation sheaths relative to expansion sheaths. The second categorization is based on speed variations across sheaths. Employing linear least squares regression, we categorize four distinct speed profiles of the sheath plasma. We find that the associated shock properties and solar cycle phase do not impact the occurrence of such variations. Our results also highlight that the properties of the driving MEs are a major source of variability in the sheath properties. Through logistic regression, we conclude that the magnetic field strength and the ME speed in the frame of the solar wind are likely drivers of these speed variations.

\end{abstract}


\section{Introduction}

\justify

Coronal mass ejections (CMEs) are manifestations of instabilities in the solar corona, leading to eruptions of magnetized plasma from the solar atmosphere \citep[e.g.][]{Green:2018}. CMEs are dynamic and expanding magnetically-dominated structures that originate from both quiescent and active regions \citep[e.g.][]{Manchester:2017} and propagate in and interact with the ambient solar wind \citep[e.g.][]{Demoulin:2009}. As the prominent carrier of prolonged intervals of southward interplanetary magnetic field \citep[IMF; e.g.][]{Farrugia:1993b}, CMEs with their enhanced internal magnetic fields are the main cause of intense geomagnetic disturbances \citep[e.g.][]{Gopalswamy:2006,Zhang:2007,Echer:2008,Richardson:2012}.

\justify

In interplanetary space, a fast CME usually consists of three separate structures: a fast-forward shock, a sheath, and a magnetic ejecta (ME) \citep[e.g.][]{Kilpua:2017b}. The sheath is defined as the region with compressed and turbulent solar wind plasma upstream of the ME. Formation of this sheath region happens because of two mechanisms: propagation of the ME through the solar wind (pure propagation sheath) and the expansion of the ME (pure expansion sheath). However, pure propagation and pure expansion sheaths are rare in terms of occurrence, since formation of most sheaths are the result of these two mechanisms acting simultaneously \citep{Siscoe:2008}. The sheath regions driven by MEs differ from planetary magnetosheaths and the heliosheath, since the \textbf{MEs also expand} \citep{Siscoe:2008,Demoulin:2009}. When the lateral deflection speed at the nose of the ME is comparable or less than the ME expansion speed, the solar wind may not be able to flow around the ME \citep{Siscoe:2008}. As a result, the solar wind entering the sheath may remain trapped therein \citep[e.g.][]{Owens:2017a}.   

\justify

Compared to the shock and ME, the sheath is the less studied structure in solar-terrestrial studies. However, for a significant fraction of the geomagnetic storms, the peak of the geomagnetic index is reached within the sheath itself \citep{Kilpua:2017b}. In rare instances, a preceding CME stuck inside the sheath of a following CME can enhance solar wind parameters and cause large geomagnetic storms \citep[see][]{Liu:2020}. Sheaths with higher dynamic pressures can couple strongly with the magnetosphere and significantly compress the dayside magnetopause \citep[e.g.][]{Lugaz:2016,Kilpua:2019b}. The sheath regions driven by MEs are also associated with higher power of magnetic fluctuations \citep[e.g.][]{Moissard:2019}. Such higher power of magnetic and dynamic pressure fluctuations can be responsible for substantial and long‐term depletions of the outer radiation belt electrons \citep[e.g.][]{Hietala:2014,Turner:2019}. Acceleration of solar energetic particles (SEPs) also has a strong reliance on the shock geometry and the structure of the sheath \citep{Manchester:2005,Lugaz:2016}. 

\justify

\textbf{However, the relevant importance of ME-driven sheaths is not only limited to its contribution in the aforementioned effects}. The layers of ME-driven sheaths contain information about the interaction of the ME with the background solar wind, since these layers are accumulated through the ME journey from the Sun \citep{Kaymaz:2006}. Sheath plasma measured near 1 AU can be composed of material with coronal origins as well \citep[see][]{Lugaz:2020a}, that can provide insight about the early phase of CME evolution. Sheath interfaces are also suitable to search for signatures of plasma discontinuities and reconnection exhausts \citep[e.g.][]{Feng:2013}. In addition, time series of solar wind plasma and magnetic field parameters corresponding to the front of the sheath can serve as a precursor for an upcoming intense southward IMF period \citep[see][]{Salman:2018}. Magnetic fluctuations within sheaths can also provide insight about the formation of coherent structures \citep{Kilpua:2020}.

\justify

For decades, CMEs have been analyzed with remote-sensing observations and in-situ measurements. Although interplanetary probes provide direct measurements of solar wind plasma and IMF \citep[e.g.][]{Richardson:2010,Jian:2006,Jian:2011,Jian:2018}, these measurements are limited to a single trajectory along the CME \citep{Russell:2002}, that make it difficult to obtain a global perspective of CME structures \citep{AlHaddad:2011,Owens:2017a,Lugaz:2018,AlaLahti:2020}. Since magnetic fields within sheaths exhibit significant variability \citep[e.g.][]{Kataoka:2005}, obtaining global configurations for sheaths become even more challenging with such one-point measurements. As a result, different statistical methods are used to understand the complexity of sheaths. A first method uses observations of the same ME-driven sheath by multiple spacecraft at different heliocentric distances that are nearly radially aligned \citep[e.g.][]{Good:2020,Lugaz:2020a,Salman:2020a}. This method sheds light on the radial growth of CME structures (sheath and ME) and decline in the magnetic field as the CME propagates away from the Sun. This method is mostly limited to case studies because of the rarity of CMEs measured in perfect radial conjunctions. In addition, small longitudinal separations between the measuring spacecraft can result in significant variations \citep[e.g.][]{Kilpua:2011,Lugaz:2018}. The relative lack of plasma measurements in the inner heliosphere also critically constrained the applicability of this method, until the recent launches of Parker Solar Probe \citep{Fox:2016} and Solar Orbiter \citep{Mueller:2013}. 

\justify

A second method consists of analyzing in-situ measurements of different sheaths at various radial distances from the Sun \citep[e.g.][]{Winslow:2015,Janvier:2019}. This method is convenient for determining the average properties at different heliocentric distances. However, such a method does not consist of measurements of the same sheath (at two or more distinct heliocentric distances) like the previous method. As a result, inherent CME-to-CME variability makes the unambiguous interpretation of radial dependency extremely difficult \citep[e.g.][]{Lugaz:2020b,Salman:2020a}.

\justify

A third method is focused on in-situ measurements of an extensive list of ME-driven sheaths near 1 AU \citep{Guo:2010,Kilpua:2013,Mitsakou:2014,Masias:2016,Rodriguez:2016,Kilpua:2017b,Jian:2018,Kilpua:2019b,Regnault:2020,Salman:2020b}. This method provides the opportunity to examine the correlation between the sheath and ME and the geoeffectiveness of CME structures in a statistical fashion.  

\justify

In this study, we follow the third approach but perform a more in-depth analysis, beyond looking at the correlation between average sheath and ME properties. We specifically focus on the sheath-to-sheath variability and how specific properties of the MEs affect the sheath properties. We carry out a comprehensive analysis of 106 ME-driven sheaths, preceded by shocks and observed by \textbf{either of} the twin STEREO \citep[Solar Terrestrial Relations Observatory;][]{Kaiser:2005} spacecraft from 2007-2016. We use in-situ measurements within sheaths to understand their complex configurations. The paper is organized as follows.  In Section~\ref{sec:DM}, we present the set of data used for this study and explain the methodology for the two different categorizations of ME-driven sheaths. Comparisons of sheaths based on these categorizations with the superposed epoch technique and statistical models are highlighted in Section~\ref{sec:Results}. In Section~\ref{sec:Dis}, we discuss our findings and conclude.  

\section{Categorization of ME-driven Sheaths} \label{sec:DM}

\subsection{Data}

\justify

The magnetic field measurements come from the magnetometer of the IMPACT \citep{Luhmann:2008} instrument and the plasma measurements come from the PLASTIC \citep{Galvin:2008} instrument, on board the twin STEREO spacecraft. We use the magnetic field data with a time resolution of 1/8 sec and proton data with a time resolution of 60 seconds.

\subsection{Methodology}

\justify

This study is a follow-up to our previous work where we investigated the generic profiles of CME structures near 1 AU \citep{Salman:2020b}. The CMEs were categorized based on the presence of shocks and sheaths. We also explored statistical relationships between sheaths preceded and not preceded by shocks, driven by MEs with comparable leading edge speeds. In the present study, we focus on the 106 ME-driven sheaths [preceded by shocks and identified as Cat-I in \citet{Salman:2020b}], selected from the STEREO CME list of \citet{Jian:2018}. This provides the opportunity for an extended analysis of ME-driven sheaths (preceded by shocks) with an exhaustive database, based on two entirely different categorization schemes. We use the sheath boundaries as listed in \citet{Jian:2018} with minor adjustments for some events based on an ``Automated Sheath Identification Algorithm" and visual assessment \citep[see][]{Salman:2020b}.

\subsubsection{Sheaths Categorized by Formation Mechanisms} \label{ssec:formation}

\justify

For the first categorization, we focus on the potential formation mechanisms of ME-driven sheaths. \textbf{These sheaths are combinations of a propagation sheath, where the solar wind is deflected sideways at the nose of the magnetic obstacle and flows around it and an expansion sheath, where the magnetic obstacle expands but does not propagate along with the background solar wind that causes plasma to pile-up upstream of the ME \citep{Siscoe:2008}}. Expansion sheaths are unique to CMEs and differ from the most commonly studied sheaths, planetary magnetosheaths, since planetary magnetosheaths are almost pure propagation sheaths \citep[e.g.][]{Kaymaz:2006,Siscoe:2008}. In an attempt to quantitatively \textbf{estimate} that either the propagation or expansion of the MEs has the significant contribution in their formations, we categorize the sheaths with three ME Mach numbers, M$_{prop}$, M$_{exp}$, and M$_{pseudo}$, based on the propagation, expansion, and leading edge speeds. The propagation speed does not include contribution from expansion. Therefore, this removes any potential overlap between M$_{prop}$ and M$_{exp}$. The reason for introducing M$_{pseudo}$ is to ensure that the sheaths of the two groups have similar M$_{pseudo}$ (i.e., similar ratios of the ME leading edge speeds in the frame of the solar wind to the characteristic speeds in the solar wind). This will ensure that the examined sheath properties of the two groups only differ due to the relative importance of their potential formation mechanisms (propagation and expansion). The three Mach numbers are defined as following:

\begin{equation}
\text{M$_{prop}$}=\frac{\text{v$_{ejecta}$ - v$_{sw}$}}{\text{v$_{ms}$}}
\end{equation}

\begin{equation}
\text{M$_{exp}$}=\frac{\text{v$_{exp}$}}{\text{v$_{ms}$}}
\end{equation}

\begin{equation}
\text{M$_{pseudo}$}=\frac{\text{v$_{le}$ - v$_{sw}$}}{\text{v$_{ms}$}} \label{eq:3}
\end{equation}

\justify

Here, v$_{ejecta}$ is the ME propagation speed, calculated as the average speed of the ME, v$_{sw}$ is the solar wind speed, v$_{ms}$ is the fast magnetosonic speed of the local solar wind, v$_{exp}$ is the ME expansion speed, calculated as half the difference between the leading and trailing edge speeds \citep[see][]{Owens:2005}, and v$_{le}$ is the ME leading edge speed. Both v$_{sw}$ and v$_{ms}$ are calculated as the average in a 2-hr time interval upstream of the shock.

\justify

We identify 28 propagation sheaths for which the M$_{exp}$$\leq$0.4 and M$_{prop}$$\geq$0.6. In a similar manner, we identify 18 expansion sheaths where the reverse is the case, i.e., M$_{prop}$$\leq$0.4 and M$_{exp}$$\geq$0.6. Although these limits are somewhat subjective, if we observe the average values (see Table~\ref{tab:ANOVA1}) for propagation sheaths (M$_{prop}$\textgreater 1 and M$_{exp}$$\sim$0) and expansion sheaths (M$_{exp}$\textgreater 1 and M$_{prop}$$\sim$0), this at least quantitatively ensures significant contribution from either propagation or expansion in forming these sheaths. Since the sheath properties are strongly correlated with the ME leading edge speeds, we impose one more criterion in categorizing these sheaths, that is 1\textless M$_{pseudo}$\textless 2. \citet{Lugaz:2017a} looked at specific cases of MEs driving shocks at 1 AU for which M$_{pseudo}$\textgreater 1 but M$_{prop}$\textless 1, i.e., where the propagation speed by itself is not sufficient to explain the formation of shocks. Here, we expand on that study by considering both propagation and expansion sheaths. We have \textbf{60} sheaths that do not fall in either category, i.e., both propagation and expansion aspects of the ME kinematics play prominent roles in their formations. This provides further confirmation that typical ME-driven sheaths are ``hybrid" sheaths, featuring aspects of both propagation and expansion \citep{Siscoe:2008}.

\subsubsection{Sheaths Categorized by Variations in Speeds} \label{ssec:Profile}

\justify

\begin{figure*}[htbp!]
  \centering
        \includegraphics[width=1.0\linewidth]{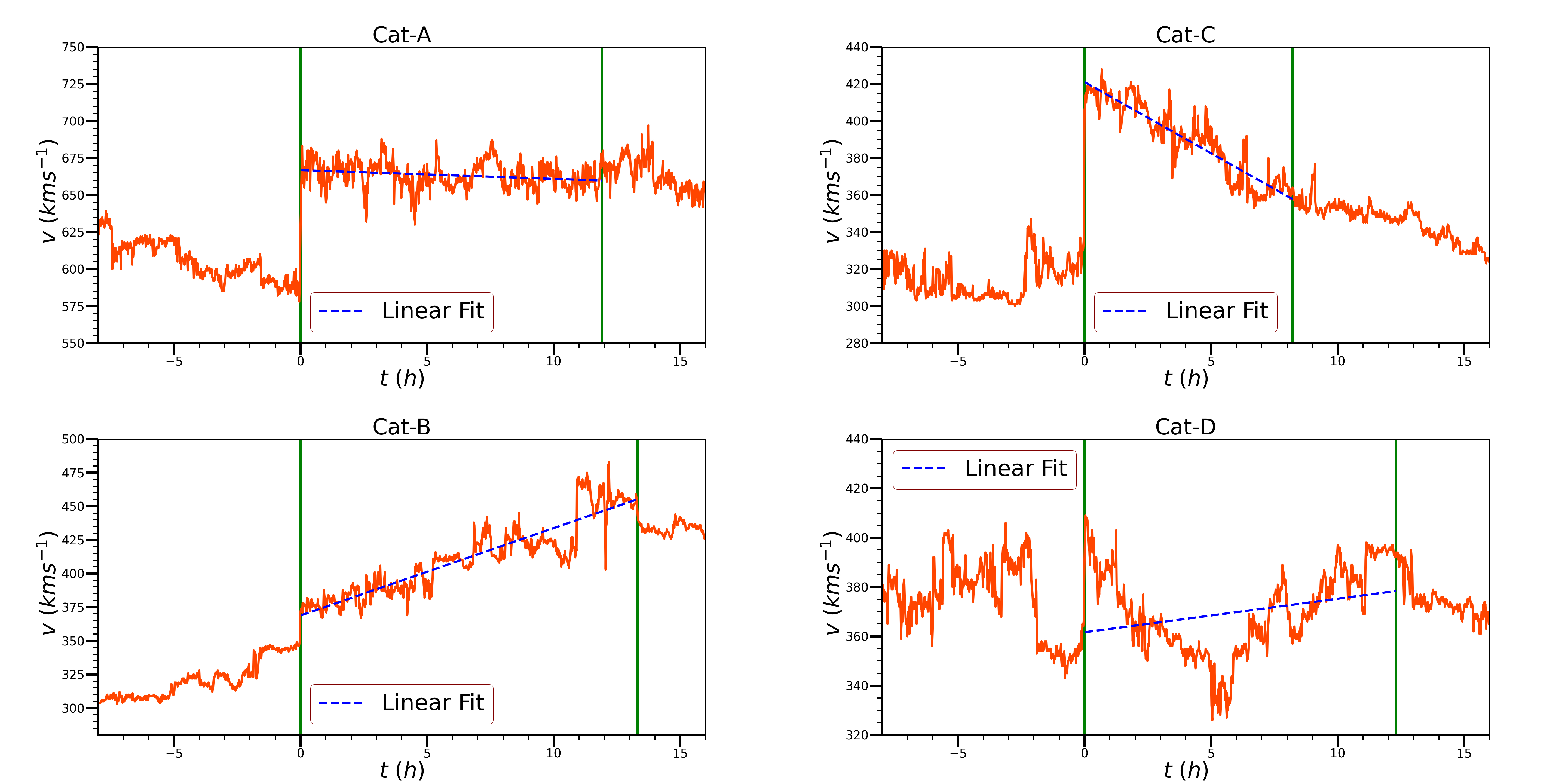}
        \caption{Speed variations within ME-driven sheath regions categorized with linear least squares regression. The vertical green solid lines bound the sheath region. The navy dashed lines represent the best linear least squares fits of the solar wind speeds within the sheath regions. \textbf{CME arrival times: Cat-A on 2014 January 29 at 5:20 UT (STEREO-B), Cat-B on 2012 October 5 at 2:51 UT (STEREO-A), Cat-C on 2011 January 17 at 15:46 UT (STEREO-B), and Cat-D on 2014 February 25 at 12:16 UT (STEREO-A).}}
         \label{fig:Categorization}
  \end{figure*}

In the second scheme, we categorize the 106 ME-driven sheaths based on their temporal variations of speeds. The rationale behind this categorization is to search for sheath characteristics (based on their speed profiles) and then try to see if we can find differences in the associated shocks and MEs, rather than looking for different driving MEs and seeing if the sheaths are different. The initial grouping relied upon the visual identification of the sheath speed profile. We categorize these sheaths into those with a constant (Category-A), increasing (Category-B), decreasing (Category-C), and complex (Category-D) speed profile, throughout the entire interval of the sheath (see Figure~\ref{fig:Categorization}). These profiles are not anomalies, with a significant number of sheaths exhibiting these trends. However, this sort of visual categorization can be subjective and introduce selection bias. Therefore, in an attempt for a more rigid categorization, we perform a linear least squares regression to identify quantifiable criteria for the categorization. We determine that the following two parameters of the regression model inherently categorize these sheaths into four distinct categories: the first parameter is the slope of the regression line relative to the average sheath speed and the second parameter is the error in the linear fitting of the sheath speed relative to the average sheath speed. The reason behind normalizing these two parameters is to minimize the influence of the sheath speed itself on the categorization.

\justify

The speed profiles for three out of the four categories exhibit marked linearities. Therefore, for the quantitative categorization of these three categories, the first criterion is that the error* (error of the regression model relative to the average sheath speed) is less than 4{\%}. The second criterion is reliant upon the slope of the regression line. For Cat-A sheaths (constant speed profiles, top left panel of Figure~\ref{fig:Categorization}), the slope* (slope relative to the average sheath speed and in units of h$^{-1}$) falls within this range: -0.4{\%} h$^{-1}$\textless slope*\textless 0.4{\%} h$^{-1}$. For Cat-B sheaths (increasing speed profiles, bottom left panel of Figure~\ref{fig:Categorization}), the slope* is greater than 0.4{\%} h$^{-1}$. For Cat-C sheaths (decreasing speed profiles, top right panel of Figure~\ref{fig:Categorization}), the slope* is negative and less than 0.4{\%} h$^{-1}$. The quantity 0.4{\%} h$^{-1}$ can be explained in this way: for a typical sheath duration of 10 hours and an average sheath speed of 500 km\,s$^{-1}$, a Cat-A sheath must have variations less than $\sim$20 km\,s$^{-1}$, whereas larger variations would be categorized as a Cat-B or Cat-C sheath. For the last category or Cat-D sheaths (complex speed profiles, bottom right panel of Figure~\ref{fig:Categorization}), we only rely on the error* for categorization, since their speed profiles do not exhibit any single definite linear trend. Therefore, the error of the regression model is expected to be significant. Cat-D sheaths therefore have an error* that is greater than 4{\%}. In total, we identify 26 Cat-A sheaths, 24 Cat-B sheaths, 20 Cat-C sheaths, and 36 Cat-D sheaths. \citet{Regnault:2020} categorized MEs into three groups based on their relative speeds, $\Delta$v (the difference between the average ME speed and the average upstream solar wind speed). The superposed epoch profiles for the speed within sheaths of their three ME groups resemble the speed profiles within sheaths of our categorization. They group the MEs based on $\Delta$v (similar to M$_{prop}$) and then look at the associated sheaths, whereas we categorize the sheaths based on their speed profiles and then try to find what causes them. In \citet{Regnault:2020}, superposed epoch profiles of sheaths, driven by MEs with low $\Delta$v, medium $\Delta$v, and high $\Delta$v show a decreasing speed profile (similar to our Cat-C sheaths), constant speed profile (similar to our Cat-A sheaths), and increasing speed profile (similar to our Cat-B sheaths) respectively.    

\section{Results} \label{sec:Results}

\subsection{Propagation versus Expansion Sheaths}

\subsubsection{Comparison of Average Properties} \label{ssec:comparison}

\justify

To examine the statistical relationships between these two types of sheaths, we perform analysis of variance (ANOVA) on the 28 sheaths identified as propagation sheaths and the 18 sheaths identified as expansion sheaths in Section~\ref{ssec:formation}. ANOVA is a parametric test that compares the means of two or more independent groups to determine statistical evidence whether or not the associated group means are significantly different. We list the results from ANOVA in Table~\ref{tab:ANOVA1}. \textbf{In Table~\ref{tab:ANOVA1}, column 1 lists the examined parameters, column 2 lists the p-values from ANOVA that provide the measure of statistical significance (with 95{\%} confidence) of a particular parameter, and column 3 and 4 list the average values for propagation sheaths and expansion sheaths respectively. In column 1, M$_{ms}$ is the shock magnetosonic Mach number, M$_{prop}$, M$_{exp}$, and M$_{pseudo}$ are the three ME Mach numbers (see Section~\ref{ssec:formation}), S$_{sheath}$ is the sheath thickness (in the radial direction), and $\theta_{nr}$ is the angle that the shock normal makes with the radial direction which can be used as an approximation of spacecraft crossing \citep[see][]{Paulson:2012}}. We use the normal vector of the shock ($\hat{\boldsymbol{n}}$) listed in the Heliospheric shock database of \citet{Kilpua:2015a} to calculate this angle. We find the following from ANOVA:

\justify

1) Both types of sheaths are driven by MEs with similar magnetic field strengths \textbf{and magnetic field strengths within propagation sheaths and expansion sheaths are also similar}.

\justify

2) Expansion sheaths are statistically larger (in the radial direction).

\justify

3) \textbf{In comparison with expansion sheaths}, propagation sheaths are statistically denser, \textbf{while the same is true for their associated ME drivers.}

\justify

4) Expansion sheaths are statistically faster.

\begin{table}[htbp]
  \centering
  \caption{ANOVA showing the variation between sample means in two categories of ME-driven sheaths. The p-values listed here represent whether the variance between the means of the two categories are significantly different. P-values representing statistical significance (\textless 0.05) are shown in bold. Note: PS=Propagation Sheath, ES=Expansion Sheath.}
    \begin{tabular}{lllll}
    \toprule
     & {ANOVA} & \multicolumn{2}{c}{Average Value} \\\hline\hline
    {Parameter} & {P-value} & {PS} & {ES} \\\hline
    B$_{ejecta}$ (nT) & 0.11 & 9.60  & 8.20 \\\hline
    N$_{ejecta}$ (cm$^{-3}$) & \textbf{0.00} & 5.80  & 3.00 \\\hline
    M$_{ms}$ & 0.19 & 1.37   & 1.17 \\\hline
    M$_{prop}$ & \textbf{0.00} & 1.20   & 0.10 \\\hline
    M$_{exp}$ & \textbf{0.00} & 0.00   & 1.00 \\\hline
    M$_{pseudo}$ & 0.83 & 1.30   & 1.20 \\\hline
    v$_{sw}$ (kms$^{-1}$)  & \textbf{0.00} & 351   & 445 \\\hline
    v$_{le}$ - v$_{sw}$ (kms$^{-1}$)  & \textbf{0.03} & 73   & 99 \\\hline
    $\theta_{nr}$ (degrees)  & 0.94 & 31.1  & 30.7 \\\hline
    S$_{sheath}$ (AU)  & \textbf{0.00} & 0.09  & 0.15 \\\hline
    B$_{sheath}$ (nT) & 0.16 & 9.80 & 8.70 \\\hline
    N$_{sheath}$ (cm$^{-3}$) & \textbf{0.00} & 16.30 & 8.40 \\\hline
    v$_{sheath}$ (km\,s$^{-1}$) & \textbf{0.01} & 421   & 531 \\\hline
    \end{tabular}%
  \label{tab:ANOVA1}%
\end{table}%

\justify

Since $\theta_{nr}$ for both types of sheaths are almost identical (see Table~\ref{tab:ANOVA1}), we emphasize that these differences are not likely byproducts of spacecraft crossings (i.e., spacecraft encountering different portions of sheaths). Characteristic features of propagation and expansion sheaths also do not seem to have any visible solar cycle dependence as we find that a major proportion (\textgreater60{\%}) of both types of sheaths occur during the maximum phase of solar cycle 24. \textbf{The upstream solar wind speed is statistically faster for expansion sheaths that can result in lower sheath densities (see Section~\ref{sec:Dis} for the influence of upstream solar wind conditions on sheath densities)}. \textbf{In addition, the leading edge speeds of MEs driving these sheaths in the frame of the solar wind (v$_{le}$ - v$_{sw}$) are statistically different and can drive some of these differences (i.e., radial thickness of sheaths)}.

\subsubsection{Superposed Epoch Analysis} \label{sssec:SEA}

\justify

In this section, we examine in more detail the average features of propagation and expansion sheaths and their associated MEs with superposed epoch analysis (SEA), also known as Chree analysis \citep{Chree:1913}. Provided that variations of distributions are random, an averaging technique like the SEA can reveal common features that are not otherwise readily detectable. However, as these representative values are reductions of the actual distributions and run the risk of being skewed due to extreme outliers, interpretations from this technique need to be done with caution. \textbf{These average features are only true for the corresponding sets of CMEs (28 CMEs with propagation sheaths and 18 CMEs with expansion sheaths) that we have analyzed}. This is also the reason we plot the median curves as well since median values tend to be less sensitive to extreme outliers \citep[see][]{Regnault:2020}.

\justify

Since the sheath and the ME durations are different for each CME, we use two characteristic epochs: the start of the sheath and the start of the ME, to rescale individual timescales to the average timescale of each category. Average durations of propagation and expansion sheaths are 8.88h and 11.60h respectively. Then, we identify the typical ME timescales based on the average sheath to ME interval ratios (0.53 for propagation sheaths and 0.35 for expansion sheaths). This leads to average ME timescales of 16.75h and 33.14h respectively for propagation and expansion sheaths. We choose a 2h time interval upstream of the shock as the background solar wind.

\justify

All parameters are binned into 75 time bins for both sheaths. These bins are then averaged to get the same number of data points within sheaths for each event. For propagation (expansion) sheaths, we impose 17 (13) time bins for the background solar wind and 142 (215) time bins for the ME. Lastly, the averaged values with the same bin number but corresponding to different CMEs are averaged to obtain the generic profiles for the background solar wind, sheath, and ME.   

 \justify

To quantify the fluctuations of the magnetic field strength \citep[e.g.][]{Masias:2016,Kilpua:2019b,Regnault:2020,Salman:2020b}, we introduce a total root-mean-square (B$_{rms}$) that is the sum of root-mean-square-deviations for each 1-min time interval (B$_{rms^*}$, as defined in Equation \ref{eq:1}) that makes up the regions (solar wind, sheath, and ME).

\begin{equation}\label{eq:1}
\text{B$_{rms^*}$}=\sqrt{\sum_{i=1}^{n} \frac{(B_{i} - \langle B \rangle)^2}{n}\\}
\end{equation}

\justify

Here, $\langle$B$\rangle$ represents the temporal average of magnetic field strength for 1-min intervals.

\justify

\begin{figure*}[htbp!]
  \centering
        \includegraphics[width=1.0\linewidth]{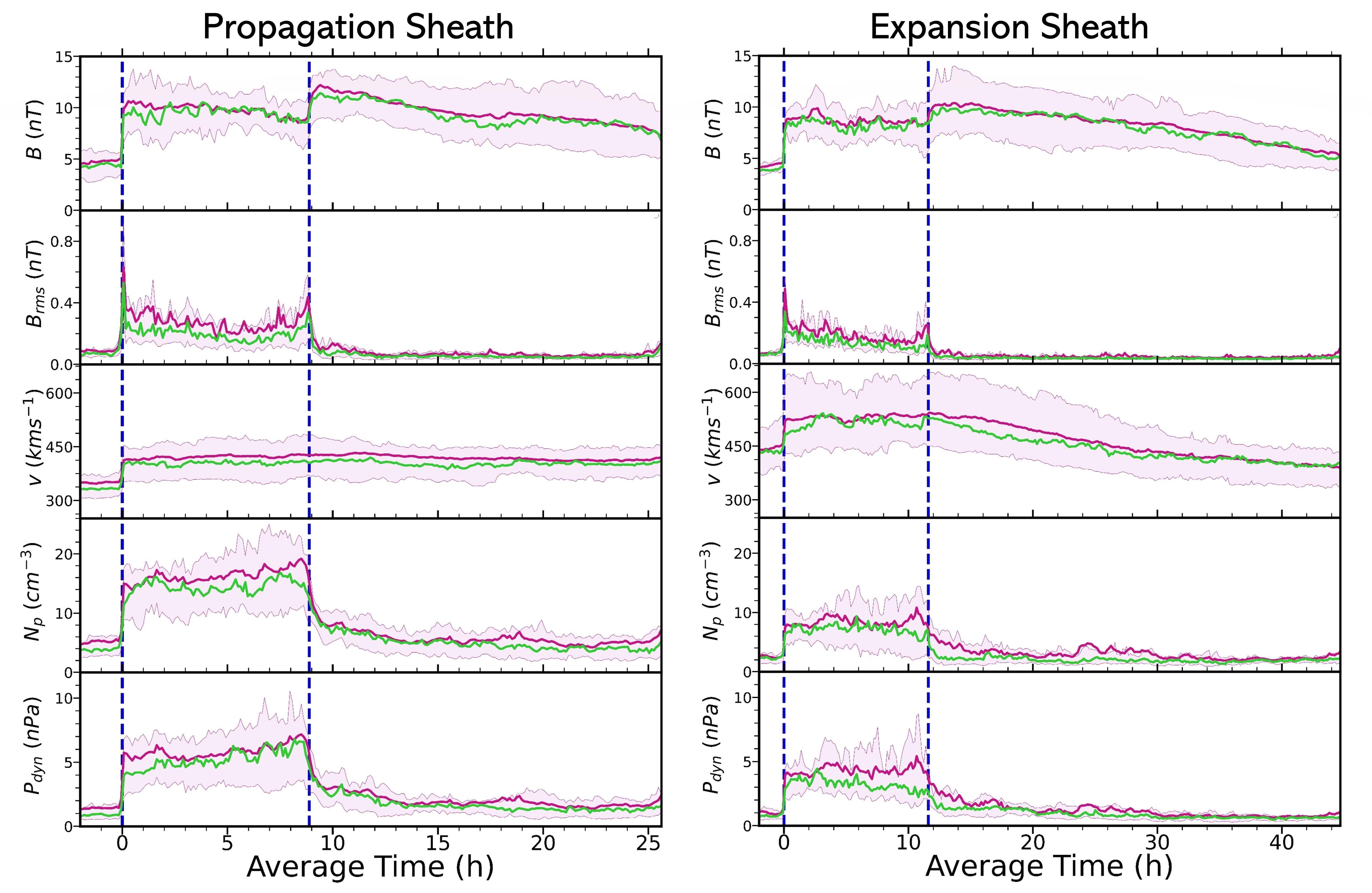}
        \caption{Superposed epoch profiles for 28 CMEs with propagation sheaths (left panels) and 18 CMEs with expansion sheaths (right panels). The panels show (from top to bottom) the magnetic field strength, total root-mean-square of magnetic field fluctuations, speed, proton density, and dynamic pressure. The purple curves show the average values, green curves show the median values, and shaded regions indicate the interquartile ranges. The vertical navy dashed lines bound the sheath region. The region to the left of the first vertical navy dashed line represents the \textbf{upstream} solar wind and the region to the right of the second vertical navy dashed line represents the ME.}
         \label{fig:SEA}
  \end{figure*}

Examining the SEA profiles for both types of sheaths (see Figure~\ref{fig:SEA}), we observe clear discontinuities for all parameters, accompanied with abrupt enhancements at the start of the sheath (shock boundary). The transitions from the solar wind to the sheath are more pronounced for propagation sheaths than expansion sheaths, except the speed. Throughout their intervals, both types of sheaths feature elevated levels of magnetic field and plasma signatures compared to the background solar wind, consistent with sheath profiles preceded by shocks \citep[see][]{Janvier:2019,Kilpua:2019b,Salman:2020b}. The density profile for the expansion sheaths shows compression near the leading edge, displaying evidence of ME expansion \citep{Owens:2020}.

\justify

Investigating the signatures of magnetic field fluctuations for both types of sheaths, we observe two prominent peaks: at the start and end of the sheath, consistent with the findings of \citet{Masias:2016,Salman:2020b} for sheaths preceded by shocks. However, the fluctuations differ in trends  for the two sheaths. For expansion sheaths, the magnetic field fluctuations correspond to a nearly uniform decrease for the entirety of the sheath, whereas the magnetic field fluctuations for propagation sheaths are associated with bi-linear trends. Inside the MEs of both types of sheaths, the magnetic field fluctuations stabilise to lower levels since MEs are low-beta structures.

\justify

The abrupt transitions in the magnetic field strength, proton density, and dynamic pressure (see first, fourth, and fifth panel of Figure~\ref{fig:SEA}) at the front of the MEs are also more pronounced for propagation sheaths. For both types of sheaths, the magnetic field strengths at the leading edges are higher compared to the rear, leading to asymmetric magnetic field profiles inside the MEs \citep[see][]{Demoulin:2008}. \citet{Demoulin:2020} suggested that these asymmetric magnetic field profiles can possibly occur due to a stronger compression on one side of the ME. For expansion sheaths, the speeds in the MEs monotonically decrease, leading to higher leading edge speeds than trailing edge speeds (see third panel on the right in Figure~\ref{fig:SEA}), in resemblance with the typical expansion undergone by MEs near 1 AU \citep{Gulisano:2010}, \textbf{whereas speeds in MEs driving propagation sheaths are constant due to minimal expansion}.

\subsection{Sheaths with Distinct Speed Variations}

\subsubsection{Solar Cycle Dependence}

\justify

\begin{figure*}[htbp!]
  \centering
        \includegraphics[width=1.0\linewidth]{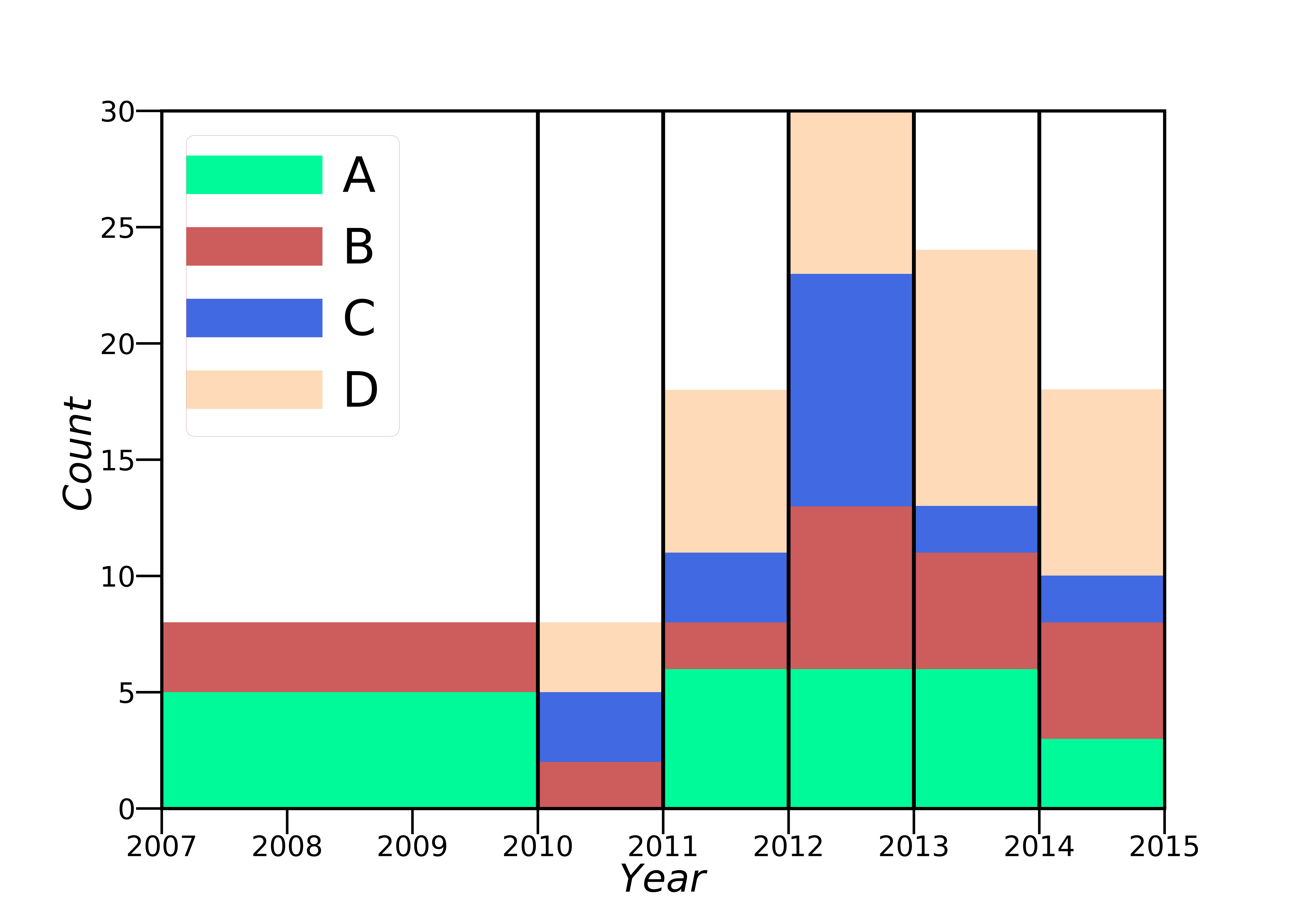}
        \caption{Solar cycle evolution of the annual occurrence rates of CMEs associated with sheaths exhibiting the four distinct speed variations.}
         \label{fig:SC}
  \end{figure*}

In this section, we examine any potential solar cycle dependence of the occurrence of CMEs with distinct speed variations in their sheaths and measured by \textbf{either of} the twin STEREO spacecraft (see Figure~\ref{fig:SC}). The time span of our study covers different phases of two solar cycles (SCs): the solar minimum of SC23 and transition to SC24 (2007-2008), the rising phase of SC24 (2009-2011), and the double-peak solar maximum of SC24 (2012-2014). Since the CME counts for the years 2007-2009 are considerably low, we combine the CME counts in this interval for a more robust statistic. The total CME count (from STEREO-A and STEREO-B) in the year 2014 has contributions from STEREO-B for the first 9 months only, as communications with STEREO-B were lost on October 1.  

\justify

During the minimum phase, we only observe the occurrence of CMEs associated with Cat-A and Cat-B sheaths. The nonexistence of CMEs associated with Cat-C and Cat-D sheaths in this time interval is somewhat unexpected. Apart from that, the occurrence rates of CMEs associated with \textbf{all} these sheaths somewhat follow similar solar cycle variations, with increasing trends coinciding with the rising phase and peaking near/at solar maximum. The occurrence rate of CMEs associated with Cat-A sheaths mostly remain constant throughout this entire time span.

\subsubsection{Superposed Epoch Analysis}

\justify

\begin{figure*}[htbp!]
  \centering
        \includegraphics[width=1.0\linewidth]{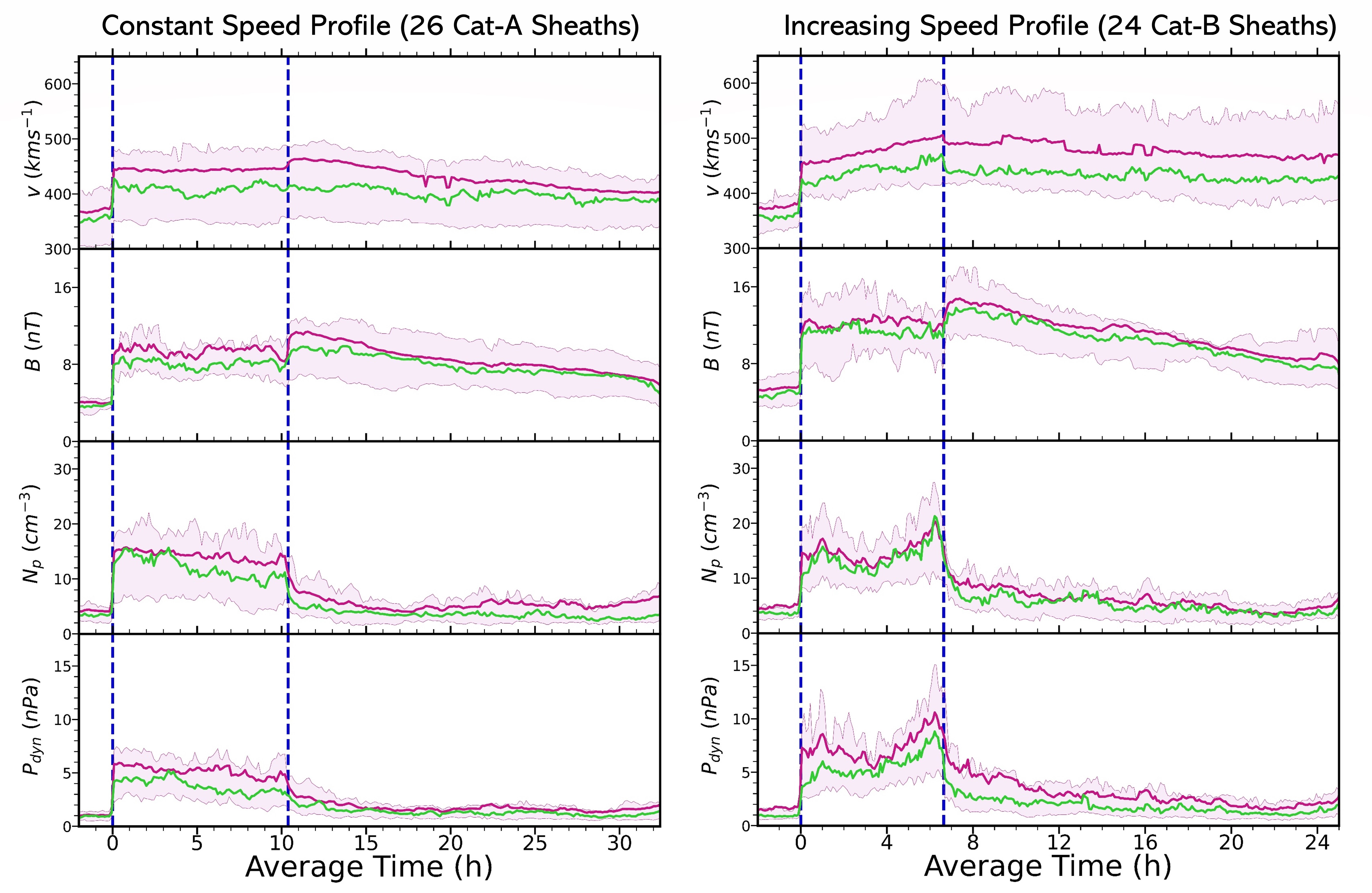}
        \caption{Superposed epoch profiles for 26 CMEs with Cat-A sheaths (left panels) and 24 CMEs with Cat-B sheaths (right panels). The panels show the speed, magnetic field strength, proton density, and dynamic pressure from top to bottom. The purple curves show the average values, green curves show the median values, and the shaded regions indicate the interquartile ranges. The vertical navy dashed lines bound the sheath region. The region to the left of the first vertical navy dashed line represents the \textbf{upstream} solar wind and the region to the right of the second vertical navy dashed line represents the ME.}
         \label{fig:SEAAB}
  \end{figure*}

\begin{figure*}[htbp!]
  \centering
        \includegraphics[width=1.0\linewidth]{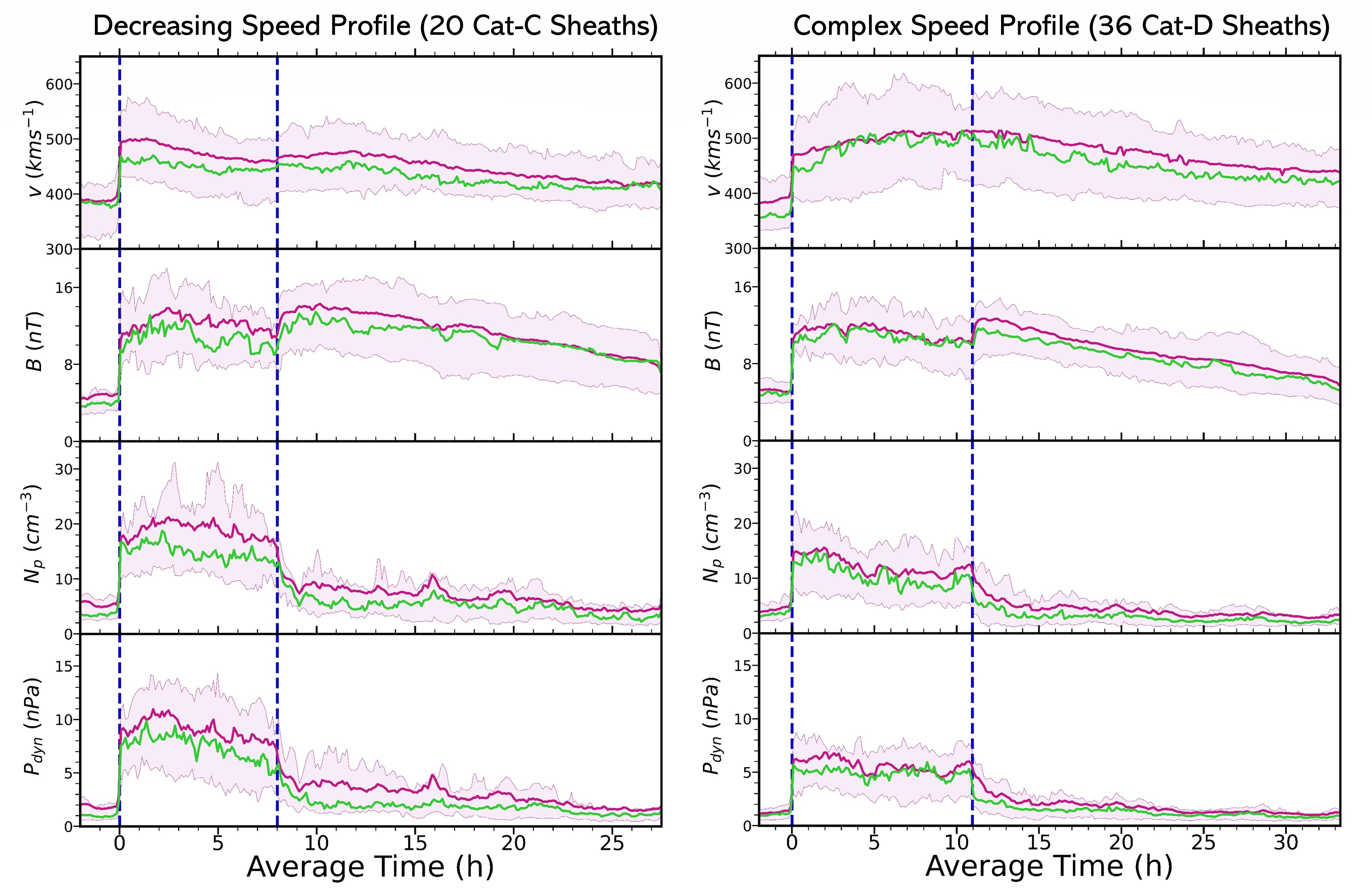}
        \caption{Superposed epoch profiles for 20 CMEs with Cat-C sheaths (left panels) and 36 CMEs with Cat-D sheaths (right panels). The panels show the speed, magnetic field strength, proton density, and dynamic pressure from top to bottom. The purple curves show the average values, green curves show the median values, and the shaded regions indicate the interquartile ranges. The vertical navy dashed lines bound the sheath region. The region to the left of the first vertical navy dashed line represents the \textbf{upstream} solar wind and the region to the right of the second vertical navy dashed line represents the ME.}
         \label{fig:SEACD}
  \end{figure*}

In a similar way to Section~\ref{sssec:SEA}, we also perform SEA for the four categories of sheaths (see Figure~\ref{fig:SEAAB} and Figure~\ref{fig:SEACD}), as categorized in Section~\ref{ssec:Profile}. The average durations of Cat-A, Cat-B, Cat-C, and Cat-D sheaths are 10.38h, 6.64h, 8.00h, and 10.95h respectively. The average sheath to ME interval ratios of 0.47, 0.36, 0.41, and 0.49 (in the same order) lead to typical ME timescales of 22.08h, 18.44h, 19.50h, and 22.35h. Here, we also have 75 time bins within sheaths. The number of time bins for the background solar wind for Cat-A, Cat-B, Cat-C, and Cat-D sheaths are 15, 23, 19, and 14 respectively and the number of time bins for the MEs are 160, 208, 183, and 153 respectively.

\justify

The observed higher speeds in the rear than in the front for Cat-B sheaths are consistent with \citet{Regnault:2020}, who found similar profiles for high $\Delta$v MEs. They suggested this occurs due to a fast ME compressing the sheath.

\justify 

\textbf{The variations in the magnetic field across the sheaths are similar for all categories}.  The magnetic field profiles of the MEs for all categories are asymmetric (i.e., stronger magnetic field at the front than in the rear). Even though Cat-A sheaths are associated with slow MEs, their MEs do not have symmetric magnetic field profiles, as found by \citet{Masias:2016} \textbf{for slow MEs}.

\justify   

\textbf{For Cat-B sheaths, we observe a density profile with two peaks, one just downstream of the shock and one close to the ME \citep[see][]{Kilpua:2017b}. Other categories have more monotonic variations of densities, with a nearly constant profile for Cat-A sheaths and decreasing profiles for Cat-C and Cat-D sheaths}. The transition from the sheath to the ME for Cat-C sheaths is not as abrupt as expected (see third and fourth panels on the left in Figure~\ref{fig:SEACD}), possibly due to magnetic erosion at the front of the ME \citep[e.g.][]{Ruffenach:2015}. The peak in density near the ME leading edge for Cat-B sheaths is an example of ``Pile-Up Compression" (PUC) region \citep[see][]{Das:2011}. This PUC region is accumulated near the Sun \citep{Kilpua:2017b}, mostly due to a larger ME expansion.

\subsubsection{Statistical Relationships in Terms of Shock Parameters} \label{sssec:Shocks}

\justify

In this section, we investigate whether the shock parameters can determine or significantly affect the properties of the following sheaths. Since the sheaths analyzed in this study are all preceded by shocks, the sheaths are composed of shocked and compressed solar wind plasma. Therefore, one could expect the sheath properties to have a strong correspondence with the properties of the preceding shocks. However, some portions of the sheath may only contain compressed (not shocked) solar wind plasma, depending on the radial distance where the shock forms \citep[see][]{Lugaz:2020a}. Similar to Section~\ref{ssec:comparison}, we perform ANOVA to identify statistically significant relationships between these four categories of sheaths, in terms of the shocks that \textbf{precede} them.

\justify

In Table~\ref{tab:ANOVA2}, B$_{down}$/B$_{up}$ and N$_{down}$/N$_{up}$ are the downstream-to-upstream ratios of the magnetic field strength and proton density respectively, v$_{shock}$ represents the shock speed in a rest frame, $\Delta$v represents the solar wind speed jump, M$_{ms}$ is the shock magnetosonic Mach number, $\theta_{Bn}$ is the angle between the shock normal and the upstream magnetic field, and $\theta_{nr}$ is the angle between the shock normal and the radial direction. \textbf{The upstream and downstream values are calculated as averages in 8-min time intervals upstream (from 9 minutes to 1 minute before the shock arrival) and downstream (from 2 minutes to 10 minutes after the shock arrival) of the shocks}. These parameters \textbf{(except $\theta_{nr}$)} are listed from the Heliospheric shock database \citep{Kilpua:2015a}. 

\justify

It is an unexpected result that the shocks of the four categories are similar in almost all measured aspects (see Table~\ref{tab:ANOVA2}) and thus shock parameters are not a useful distinguishing feature for these sheaths. However, ANOVA reveals that some of these speed variations can be possible manifestations of the nature of spacecraft measurements (characterized by $\theta_{nr}$), depending on the portion of the sheath encountered by the measuring spacecraft. As this is the only relationship that we find to be statistically significant (p-value=0.01), we run Tukey's honestly significant difference (HSD) post hoc test. Tukey's HSD post hoc test compares all possible pairs of means and controls the experimentwise error rate ($\alpha$=0.05). From this test, we see that sheaths with linear trends of speed variations (Cat-A, Cat-B, and Cat-C) are not statistically different from one another, in terms of spacecraft measurements (the three p-values for pairwise comparisons are \textgreater0.9). This test also shows that only two pairs have statistically significant differences: i) Cat-B and Cat-D, with p-value=0.03 and ii) Cat-C and Cat-D, with p-value=0.04.

\begin{table}[htbp]
  \centering
  \caption{Average values of selected parameters associated with shocks preceding the four categories of ME-driven sheaths. The p-values listed here represent whether the variance between the means of the four categories are significantly different (from ANOVA). P-values representing statistical significance (\textless 0.05) are shown in bold.}
    \begin{tabular}{lccccc}
    & \multicolumn{4}{c}{Average Value} &  ANOVA \\\hline\hline
    \multicolumn{1}{l}{Parameter} & Cat-A & Cat-B & Cat-C & Cat-D & P-value \\\hline
    B$_{down}$/B$_{up}$ & 2.31   & 2.14  & 2.55  & 2.44   & 0.59 \\\hline
    N$_{down}$/N$_{up}$ & 3.96   & 3.02     & 3.72   & 3.75   & 0.29 \\\hline
    v$_{shock}$ (km\,s$^{-1}$) & 423   & 464   & 478   & 402   & 0.10 \\\hline
    $\Delta$v (km\,s$^{-1}$) & 76  & 75  & 98  & 77  & 0.45 \\\hline
    M$_{ms}$ & 1.35  & 1.55  & 1.72  & 1.40  & 0.37 \\\hline
    $\theta_{Bn}$ (degrees) & 57  & 67  & 60  & 64  & 0.53 \\\hline
    $\theta_{nr}$ (degrees) & 26  & 24  & 24  & 37  & \textbf{0.01} \\\hline
    \end{tabular}%
  \label{tab:ANOVA2}%
\end{table}%

\justify

\textbf{As Cat-A, Cat-B, and Cat-C sheaths are not statistically different for $\theta_{nr}$, the distances from the noses of the shocks at which the spacecraft crossings occur are presumably not the primary drivers of the differences described in this section. In this sense, it differs from the analysis of \citet{Paulson:2012} which attributed Cat-B sheaths (increasing speed profiles) to the possibility of crossings away from the noses. Only Cat-D sheaths, which are sheaths that do not fit with linear trends of speeds may be associated with spacecraft crossings farther away from the noses of the shocks (as is clear from the larger $\theta_{nr}$). This analysis also points towards the fact that complex speed profiles within sheaths (Cat-D) are more likely for crossings at larger distances from the noses of the shocks.}

\subsubsection{Statistical Differences in Terms of Sheath and ME Parameters}

\justify

In this section, we aim to identify whether the speed variations within sheaths are associated with intrinsic differences of the CME structures. As shown in Table~\ref{tab:ANOVA3}, we find that there are a number of statistically significant differences between the four categories. ANOVA reveals that:

\justify

1) The magnetic field strength and density of the MEs have statistically significant differences between the categories. \textbf{For both these parameters, we run Tukey's HSD post hoc test for pairwise comparisons. We only find the density of the MEs associated with Cat-C and Cat-D sheaths to be statistically different from one another (with p-value=0.04)}. 

\justify

2) The ME propagation and expansion speeds are similar between the categories. Therefore, these sheaths exhibit such distinct trends of variations, despite the driving MEs propagating with similar average and expansion speeds.

\justify

3) The radial thickness and density of the four categories of sheaths are statistically different. \textbf{With Tukey's HSD post hoc test, we find that the radial thickness of Cat-B and Cat-D sheaths (with p-value=0.00) and density within Cat-C and Cat-D sheaths (with p-value=0.02) are statistically different from one another}.

\justify

\textbf{The findings presented in this section point towards the fact that sheaths with linear trends of speeds are statistically similar to one another. The statistical differences only exist between sheaths with non-linear trends of speeds (Cat-D) and sheaths with linear trends of speeds (Cat-B and Cat-C).}  

\begin{table}[htbp]
  \centering
  \caption{Average values of selected sheath and ME parameters associated with the four categories of ME-driven sheaths. The p-values listed here represent whether the variance between the means of the four categories are significantly different (from ANOVA). P-values representing statistical significance (\textless 0.05) are shown in bold.}
    \begin{tabular}{lccccc}
    & \multicolumn{4}{c}{Average Value} &  ANOVA \\\hline\hline
    \multicolumn{1}{l}{Parameter} & Cat-A & Cat-B & Cat-C & Cat-D & P-value \\\hline
    B$_{ejecta}$ (nT) & 8.5   & 11.2  & 11.5  & 9.2   & \textbf{0.02} \\\hline
    N$_{ejecta}$ (cm$^{-3}$) & 5.5   & 6     & 6.8   & 4.2   & \textbf{0.04} \\\hline
    v$_{ejecta}$ (km\,s$^{-1}$) & 429   & 482   & 446   & 475   & 0.27 \\\hline
    M$_{pseudo}$ & 1.45  & 1.94  & 1.33  & 2.09  & 0.07 \\\hline
    v$_{exp}$ (km\,s$^{-1}$) & 25    & 14    & 21    & 34    & 0.11 \\\hline
    S$_{sheath}$ (AU) & 0.11  & 0.08  & 0.09  & 0.13  & \textbf{0.00} \\\hline
    B$_{sheath}$ (nT) & 9.4   & 12.3  & 12.3  & 11.1  & 0.10 \\\hline
    N$_{sheath}$ (cm$^{-3}$) & 14.2  & 14.8  & 18.7  & 12.4  & \textbf{0.04} \\\hline
    v$_{sheath}$ (km\,s$^{-1}$) & 444   & 478   & 477   & 500   & 0.30 \\\hline
    \end{tabular}%
  \label{tab:ANOVA3}%
\end{table}%

\subsubsection{Influence of ME Parameters in Driving Distinct Speed Variations within Sheaths} \label{sssec:LR}

\justify

In this section, we measure the significance of the four ME parameters with the lowest p-values (\textbf{B$_{ejecta}$, N$_{ejecta}$, M$_{pseudo}$, and v$_{exp}$}, see Table~\ref{tab:ANOVA3}) for a sheath to be identified as a specific category, with logistic regression (LR). LR is a predictive linear algorithm and explains the relationship between a dependent (or response) variable that is dichotomous in nature and one or more independent (or predictor) variables. Therefore, LR can be used for classification purposes. In our case, we use the LR analysis to assess the dependency of the binary outcome of the response variable (whether a sheath will be identified as a specific category or not) on the predictor variables (the four ME parameters, see Table~\ref{tab:LR}), in a similar manner to \citet{Riley:2012}. They also used LR analysis to identify the likelihood for a CME to be classified as a MC or not based on a number of predictor variables.

\justify

Table~\ref{tab:LR} summarizes the LR analysis. The estimated regression coefficients (column 2 in Table~\ref{tab:LR}) for the predictor variables represent the effect of a unit change in a predictor variable (when the other predictor variables are held constant) in the log odds of the response variable, with 95{\%} confidence. The sign of a coefficient represents the direction of the relationship between a predictor variable and the response variable. A positive coefficient will generally indicate that with the increase in a predictor variable (when the other predictor variables in the model do not change), the event is more likely to happen. The p-values (column 5 in Table~\ref{tab:LR}) provide the measure of statistical significance (with 95{\%} confidence) of a particular ME parameter. The intercepts in Table~\ref{tab:LR} represent the log of the odds of the response variables when all the predictor variables are 0.  

\justify

From the LR analysis, we infer that the following ME parameters have marked influences on the categorization: B$_{ejecta}$, N$_{ejecta}$, and M$_{pseudo}$. This may give clues as to which physical processes cause such speed variations. However, it is important to note that LR only provides statistical evidence, since higher p-values do not mean that a particular parameter (e.g. ME expansion speed) is not significant. It only represents the absence of statistical evidence or undetected evidence by the statistical technique used \citep[see][]{Riley:2012}. From the LR analysis, the likelihood of the categorization depends on:    

\justify

a) MEs with strong magnetic fields are less likely to drive Cat-A sheaths.
 
\justify

b) None of these parameters have any statistical influence for a sheath to be categorized as Cat-B.

\begin{table}[htbp]
  \centering
  \caption{Logistic regression models for categorization of ME-driven sheaths. Column one lists the predictor variables, column two gives the estimated value of the regression coefficient for each predictor variable, column three gives the standard errors for these estimates, column four gives the results of the Chi-squared test, and column five lists the p-values. P-values representing statistical significance (\textless 0.05) are shown in bold.}
    \begin{tabular}{lcccc}
    Parameter & Estimate & Std. Error & Chi-Square & P-value \\\hline\hline
    \multicolumn{5}{c}{\textit{\textbf{For log odds of Cat-A/Not Cat-A}}} \\\hline
    Intercept & 0.3502 & 0.765 & 0.21  & 0.6471 \\\hline
    B$_{ejecta}$ (nT) & -0.1691 & 0.0858 & 3.88  & \textbf{0.0489} \\\hline
    N$_{ejecta}$ (cm$^{-3}$) & 0.0936 & 0.0781 & 1.44  & 0.2308 \\\hline
    M$_{pseudo}$ & -0.2446 & 0.2414 & 1.03  & 0.3109 \\\hline
    v$_{exp}$ (km\,s$^{-1}$) & 0.0003 & 0.0068 & 0.00     & 0.9600 \\\hline
    \multicolumn{5}{c}{\textit{\textbf{For log odds of Cat-B/Not Cat-B}}} \\\hline
    Intercept & -2.1423 & 0.7413 & 8.35  & 0.0039 \\\hline
    B$_{ejecta}$ (nT) & 0.0828 & 0.0639 & 1.68  & 0.1948 \\\hline
    N$_{ejecta}$ (cm$^{-3}$) & 0.0028 & 0.0677 & 0.00     & 0.9666 \\\hline
    M$_{pseudo}$ & 0.1487 & 0.2081 & 0.51  & 0.4749 \\\hline
    v$_{exp}$ (km\,s$^{-1}$) & -0.0134 & 0.0082 & 2.66  & 0.1029 \\\hline
    \multicolumn{5}{c}{\textit{\textbf{For log odds of Cat-C/Not Cat-C}}} \\\hline
    Intercept & -2.8609 & 0.8585 & 11.1  & 0.0009 \\\hline
    B$_{ejecta}$ (nT) & 0.1802 & 0.0784 & 5.28  & \textbf{0.0215} \\\hline
    N$_{ejecta}$ (cm$^{-3}$) & 0.1624 & 0.0818 & 3.95  & \textbf{0.0470} \\\hline
    M$_{pseudo}$ & -1.0790 & 0.4091 & 6.96  & \textbf{0.0084} \\\hline
    v$_{exp}$ (km\,s$^{-1}$) & 0.0083 & 0.0086 & 0.95  & 0.3296 \\\hline
    \multicolumn{5}{c}{\textit{\textbf{For log odds of Cat-D/Not Cat-D}}} \\\hline
    Intercept & -0.0426 & 0.7140 & 0.00     & 0.9524 \\\hline
    B$_{ejecta}$ (nT) & -0.0701 & 0.0646 & 1.18  & 0.2781 \\\hline
    N$_{ejecta}$ (cm$^{-3}$) & -0.1900 & 0.0874 & 4.72  & \textbf{0.0297} \\\hline
    M$_{pseudo}$ & 0.5164 & 0.2063 & 6.27  & \textbf{0.0123} \\\hline
    v$_{exp}$ (km\,s$^{-1}$) & 0.0037 & 0.0063 & 0.34  & 0.5587 \\\hline
    \end{tabular}%
  \label{tab:LR}%
\end{table}%

\justify

c) MEs with strong magnetic fields and high densities are likely to drive Cat-C sheaths. In addition, MEs with high Mach numbers (M$_{pseudo}$) are less likely to drive Cat-C sheaths. 

\justify

d) MEs with high Mach numbers (M$_{pseudo}$) are likely to drive Cat-D sheaths. In addition, MEs with high densities are less likely to drive Cat-D sheaths.  

\justify

To summarize, LR analysis suggests that the categorization is \textbf{considerably} coupled to the magnetic field strengths and the leading edge speeds in the frame of the solar wind (computed as M$_{pseudo}$) of the associated MEs. The reason for not including N$_{ejecta}$ here is that even when this parameter is statistically significant (in the categorization of Cat-C and Cat-D sheaths), it is still the least influential among the other parameters (higher p-values).  

\subsubsection{Comparison of Statistical Multivariate Method with in-situ Observations}

\justify

\begin{figure*}[htbp!]
  \centering
        \includegraphics[width=1.0\linewidth]{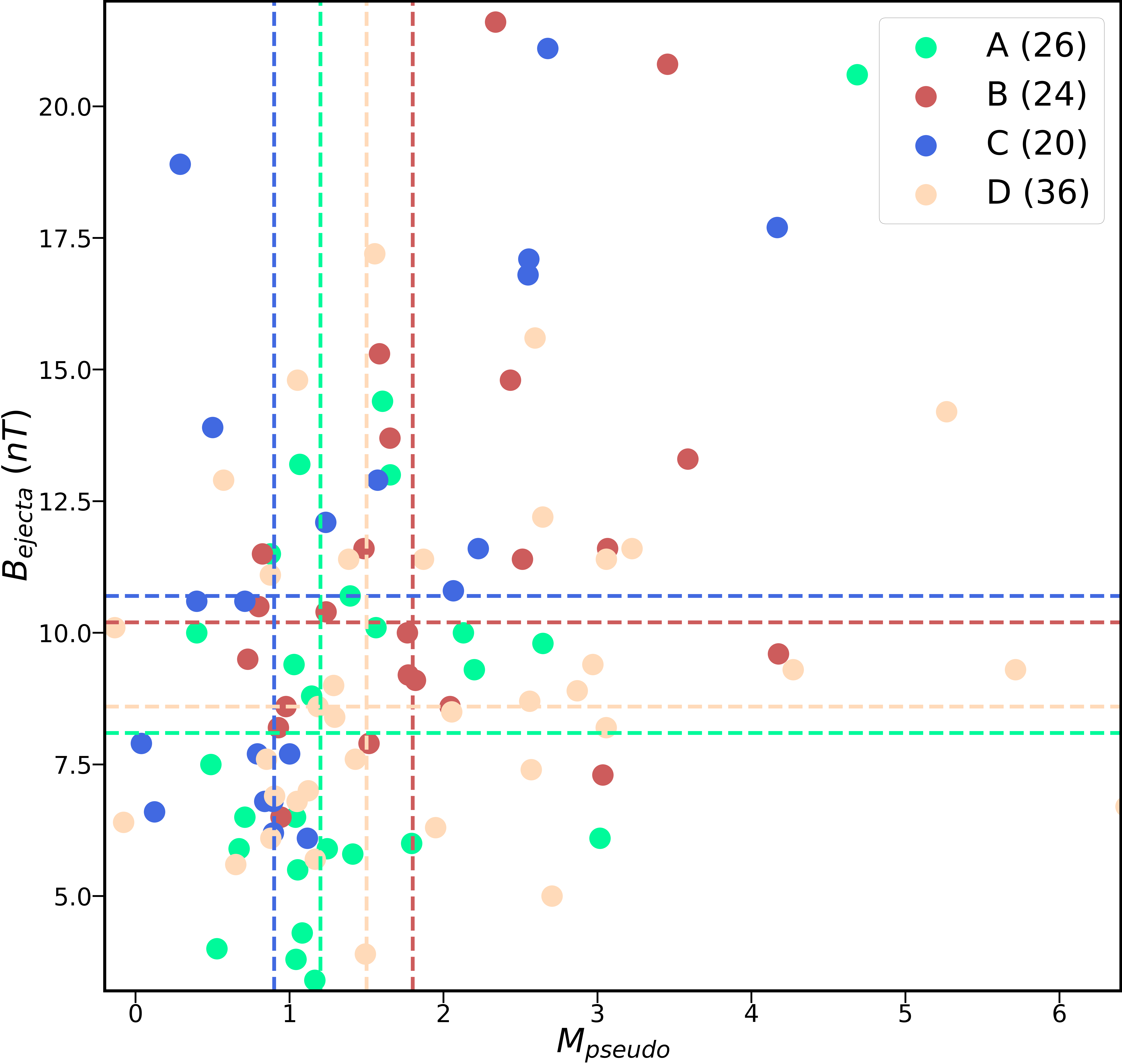}
        \caption{Distribution of the 106 CMEs (driving both shocks and sheaths near 1 AU) measured by \textbf{either of} the twin STEREO spacecraft from 2007-2016, as a function of the magnetic field strength and Mach number (M$_{pseudo}$) of the associated MEs. The CMEs are color-coded according to the four sheath categories. The horizontal and vertical dashed lines represent the \textbf{medians} of the magnetic field strength and Mach number (M$_{pseudo}$) of each category.}
         \label{fig:Scatter}
  \end{figure*}

In the previous section, we have used LR to assess the statistical influences of four ME parameters on our categorization. However, the LR analysis relied on our categorization in the first place. Therefore, it is important to identify whether the parameters M$_{pseudo}$ and B$_{ejecta}$ can distinguish the CMEs from one another or the predictions from the LR analysis are simply outcomes of selection bias. To address this, we present the four categories of sheaths as a function of the magnetic field strengths and Mach numbers (M$_{pseudo}$) of the associated MEs (see Figure~\ref{fig:Scatter}). Although the plot does not fully separate the four sheath categories, with considerable overlaps between the distributions, the trends (represented with horizontal and vertical dashed lines in Figure~\ref{fig:Scatter}) of the categories are noticeable. This plot demonstrates that statistical predictions mirror the actual distributions well, with an apparent dependence on M$_{pseudo}$ and B$_{ejecta}$ for the categorization. We observe the following:  

\justify

a) Cat-A sheaths are likely to be driven by MEs with a \textbf{moderate} M$_{pseudo}$ and weaker magnetic fields.

\justify

b) Cat-B sheaths are likely to be driven by MEs with a higher M$_{pseudo}$ and stronger magnetic fields.

\justify

c) Cat-C sheaths are likely to be driven by MEs with a lower M$_{pseudo}$ and stronger magnetic fields.

\justify

d) Cat-D sheaths are likely to be driven by MEs with a \textbf{moderate} M$_{pseudo}$ and weaker magnetic fields.

\section{Discussion and Conclusions} \label{sec:Dis}

\justify

In this study, we present an analysis of 106 ME-driven sheaths, using measurements from the STEREO probes between 2007-2016. We use a concise version of our CME catalog \citep[see][]{Salman:2020b}, focusing solely on sheaths that are preceded by shocks near 1 AU. We categorize these sheaths based on their potential formation mechanisms and observed speed variations. \textbf{We perform ANOVA to identify statistically significant differences between sheaths for both categorizations}. We \textbf{also} use SEA to extract the generic features of propagation and expansion sheaths and also the four categories of sheaths with distinct speed variations. We then use LR to examine the correspondence of these speed variations with selected parameters of the associated MEs.

\justify

The first categorization is based on how ME-driven sheaths form. We aim to quantitatively ensure (with predefined constraints on M$_{prop}$, M$_{exp}$, and M$_{pseudo}$) that the two categories of sheaths (propagation and expansion sheaths) only include sheaths for which either the propagation or the expansion of the MEs is the dominant mechanism behind their formations.

\justify

We then attempt to identify statistical differences between propagation and expansion sheaths with ANOVA. We find that both types of sheaths are driven by MEs with similar magnetic field strengths. We observe that expansion sheaths are statistically faster and thicker (in the radial direction). However, the sheath to ME interval ratio is considerably smaller for expansion sheaths compared to propagation sheaths (0.35 to 0.53). This is consistent with \citet{Siscoe:2008} who found the shock standoff distance, normalized to the radius of curvature of the expanding ME for a fast and pure expansion sheath to be less than a propagation sheath. For these expansion sheaths, the solar wind plasma does not only stack up in the radial direction. Due to the ME expansion, accreted material close to the sun can slide in the lateral direction as well, as discussed in \citet{Siscoe:2008}, therefore reducing the radial thickness. This effect should also depend on how the lateral expansion is related to the radial expansion, since the lateral expansion may happen faster than the radial expansion \citep[e.g.][]{Nieves:2018}. \citet{Gopalswamy:2006} also pointed out that based on the specific path an observing spacecraft takes through a MC (see his Figure 2), a spacecraft may encounter a thicker sheath far off the nose of the MC and a thinner sheath at the nose. Since there is no statistical difference of the shock normal angles between the propagation and expansion sheaths with respect to the radial direction (proxy for the distance of the spacecraft crossing from the shock nose, see Table~\ref{tab:ANOVA1}), their difference in the radial thickness is \textbf{presumably} not due to spacecraft crossings.

\justify

We also find that propagation sheaths are statistically denser (see Table~\ref{tab:ANOVA1}), but \textbf{the} faster propagation and slower expansion speeds of \textbf{their} MEs would be expected to result in less pile-up of solar wind material in front of it. However, their statistically lower ME speeds in the frame of the solar wind can reduce the deflection speeds at the nose of the MEs, resulting in denser sheaths. In addition, sheath density strongly relies on the upstream solar wind conditions \citep{Temmer:2021}. MEs driving propagation sheaths also propagate in denser and slower solar wind. This can also result in higher sheath densities, since the amount of solar wind material upstream of the ME would be considerably larger \citep{Temmer:2021}. \textbf{Apart from this, MEs driving propagation sheaths are subjected to less amount of drag due to their lower ME speeds in the frame of the solar wind \citep[see][]{Vrsnak:2013}. This can result in minimal variations in ME speeds (see third panel on the left in Figure~\ref{fig:SEA}).}

\justify

Then, we examine the characteristic signatures of propagation and expansion sheaths and their associated MEs with SEA. Mostly all the parameters are more elevated in propagation sheaths. The magnetic fluctuations in these sheaths follow different trends. Similar to the transition from the background solar wind to the sheath, the transition from the sheath to the ME is more abrupt (in magnetic field strength, proton density, and dynamic pressure, see Figure~\ref{fig:SEA}) for propagation sheaths.

\justify

For our second categorization, we look beyond the observed magnetic fluctuations within sheaths. This categorization is based on the speed variations observed within ME-driven sheaths (see Figure~\ref{fig:Categorization}). It builds upon an initial grouping of sheaths with visual assessment. We then use linear least squares regression to quantify the linearities in the speed variations and categorize them into four categories, based on two modified model parameters (slope* and error*).

\justify

We also investigate the solar cycle dependency for the occurrence of these distinct speed variations (see Figure~\ref{fig:SC}). As these sheaths are driven by MEs that also drive shocks near 1 AU, we find their occurrences to peak near the solar maximum of SC24. We also observe that Cat-A sheaths (driven by MEs with moderate speeds and weak magnetic fields) occur at a relatively constant rate. They could be associated with streamer blowouts that occur throughout the different phases of the solar cycle \citep{Vourlidas:2018}. It is also evident that the occurrences of Cat-C and Cat-D sheaths increase with solar activity. During solar maximum, CMEs are more likely to erupt from active regions \citep{Manchester:2017}. This can possibly suggest that Cat-C and Cat-D sheaths are likely to be driven by MEs originating from active regions on the Sun.

\justify

From the SEA, in the four categories of sheaths, the plasma and magnetic field parameters are more enhanced than the background solar wind. As expected, Cat-B sheaths have higher speeds in the rear than in the front. Cat-B sheaths also show evidences of larger ME expansion close to the Sun, with a PUC region near the ME leading edge (see third panel on the right in Figure~\ref{fig:SEAAB}).

\justify

Then, we examine possible correlations between the shock parameters and these speed variations within sheaths (see Table~\ref{tab:ANOVA2}). If the micro scale processes in the vicinity of the shocks are responsible for these distinct speed variations, it is natural to expect the properties of the shocks, associated with the four categories to be significantly different from one another. However, ANOVA results indicate that these shocks are similar in almost every way possible. One possible explanation for ME-driven sheaths exhibiting such speed profiles can be that these are manifestations of the nature of spacecraft measurements \citep[e.g.][]{Paulson:2012}. Therefore, we examine the importance of the spacecraft's trajectory through the CME, characterized by the parameter $\theta_{nr}$ (i.e., the angle between the shock normal and the radial direction). From ANOVA and the subsequent Tukey's HSD post hoc test, we find no statistical difference in the average values of $\theta_{nr}$ between the three categories with linear trends in speeds (Cat-A, Cat-B, and Cat-C). In addition, statistical differences only exist between the values of $\theta_{nr}$ for sheaths with monotonic and complex speed profiles (see Section~\ref{sssec:Shocks}). Therefore, sheaths associated with complex speed profiles (Cat-D) are more likely to be associated with shocks \textbf{measured} away from the nose (larger $\theta_{nr}$). In comparing the properties of the sheaths and the MEs with ANOVA (see Table~\ref{tab:ANOVA3}), we find that the magnetic field strength and density of the MEs and the radial thickness and density of the sheaths for the four categories are statistically different. \textbf{However, Tukey's HSD post hoc tests reveal that most of the differences only exist between sheaths with non-linear trends of speeds (Cat-D) and sheaths with linear trends of speeds (Cat-B and Cat-C)}.

\justify

After that, we examine the possibility of these speed variations being driven by intrinsic ME parameters. We use four LR models, each time examining the statistical influences of four ME parameters (B$_{ejecta}$, N$_{ejecta}$, M$_{pseudo}$, and v$_{exp}$) on a binary outcome in Section~\ref{sssec:LR}, similarly to \citet{Riley:2012}. LR reveals that the magnetic field strength and Mach number (M$_{pseudo}$) of the associated MEs are likely to be the two most important ME parameters driving these variations (see Table~\ref{tab:LR}). The statistical importance of the ME magnetic field strength provides hint that these speed variations can be attributed to the evolutionary processes within the CME. The importance of the ME magnetic field strength might also be an indirect measure of expansion, although \citet{Lugaz:2020b} found no correlation between the magnetic field near 1 AU and expansion, as measured by the decline of the magnetic field between MESSENGER/\textit{Venus Express} and 1 AU. \citet{Vrsnak:2019} also reported discrepancies between the decline of the magnetic field and rate of expansion. The other important parameter M$_{pseudo}$ has contributions from the leading edge speed (see Equation \ref{eq:3}). The leading edge speed is the sum of the propagation speed and expansion speed of the ME. From LR, we find that the ME expansion speed does not seem to have an influence on the categorization. This is somewhat surprising and the possible reason may be that too many parameters vary concurrently for each ME and our sample size is limited ($\sim$100 events). Therefore, the statistical significance of M$_{pseudo}$ indicates that the propagation speed of the ME in the frame of the solar wind is important. This is consistent with \citet{Regnault:2020} who found similar speed variations within sheaths based on this characteristic.   

\justify

We also examine the distributions of the CMEs in the M$_{pseudo}$ - B$_{ejecta}$ space (see Figure~\ref{fig:Scatter}), the two parameters LR predicts to be the most influential (statistically) in driving these speed variations. Cat-A sheaths (constant speed profile) tend to be driven by moderately slow MEs with weak magnetic fields and with a near-constant solar cycle dependency. One possibility is that they are associated with streamer blowouts, which would be consistent with all these properties. This would need to be confirmed by studying the coronal association with CMEs and potentially ion charge states. Constant speed profile would mean that the combination of ME expansion and its deceleration due to the ``magnetohydrodynamic drag" \citep[e.g.][]{Vrsnak:2013} somewhat balances out. This can weaken the broadening of the ME body and make all elements within the CME move with uniform speed (see first panel on the left in Figure~\ref{fig:SEAAB}). 

\justify

Cat-B sheaths (increasing speed profile) tend to be driven by MEs with a high Mach number (M$_{pseudo}$) and strong magnetic fields. They might be for those that have been expanding strongly in the past and the sheaths were more compressed in the innermost heliosphere (evident by the PUC region near the ME leading edge, see third panel on the right in Figure~\ref{fig:SEAAB}). With two simulated sheaths, \citet{Das:2011} pointed the significance of a larger expansion speed for a higher PUC in the lower corona. \citet{Paulson:2012} suggested such increasing speed profiles within sheaths can possibly occur due to spacecraft crossings, where it may appear that the MEs are overtaking the shocks that they are driving. They argued that if the shock normal forms a significant angle with the radial direction, the spacecraft measuring the flank of the shock will observe the shock to propagate slowly than the following ME, possibly leading to an increasing speed profile within its sheath. This should require $\theta_{nr}$ to be significantly larger for shocks associated with Cat-B sheaths. However, from the ANOVA, examining the average values (see Table~\ref{tab:ANOVA2}), we do not observe such trends. 

\justify

Cat-C sheaths (decreasing speed profile) tend to be driven by slow MEs with strong magnetic fields. They might be associated with MEs for which the expansion speed becomes an increasingly important contribution to the leading edge speed. \citet{Regnault:2020} suggested that such expansion like profiles within sheaths can occur when the ME propagates with a speed comparable to the background solar wind. This provides the sheath with more time to adjust to the surrounding conditions and the sheath expands in a similar manner to the ME.

\justify

Cat-D sheaths (complex speed profile) need to be investigated into further details. These sheaths are driven by moderately fast MEs (on average). Interestingly, $\sim$10 of the 36 Cat-D sheaths exhibit clear bi-linear trends in their speed profiles, potentially indicating a combination of Cat-B and Cat-C sheath properties. Extended analysis of them is left for a follow-up study.

\justify

This statistical study, based on observational data from in-situ instruments highlights the sheath-to-sheath variability observed near 1 AU. Our analysis allows us to derive overall statistical trends for two different categorizations of ME-driven sheath regions. However, simple geometry assumptions and quantitative limitations associated with the analysis suggest that interpretations of statistical results should be done with caution. Further investigation of sheath characteristics with multi-spacecraft observations and three-dimensional magnetohydrodynamic simulations can provide key insight regarding the complexities of CME propagation and interaction with the background solar wind, related to the observed variability within sheaths near 1 AU.

\begin{acknowledgements}

\justify

T.~M.~S and N.~L acknowledge support from NASA grants 80NSSC20K0700, 80NSSC20K0197, 80NSSC17K0556, and 80NSSC20K0431. We also acknowledge the use of Heliospheric shock database, generated and maintained at the University of Helsinki and can be found at \url{http://ipshocks.fi/}. We are grateful to the STEREO mission team and NASA/GSFC's Space Physics Data Facility's CDAWeb service (available at \url{https://cdaweb.gsfc.nasa.gov/index.html/}) for providing the data needed for this study. We also thank the anonymous reviewer for critically reading the paper and suggesting substantial improvements.

\end{acknowledgements}

\newpage

\bibliography{Salman}{}

\begin{thebibliography}{}
\expandafter\ifx\csname natexlab\endcsname\relax\def\natexlab#1{#1}\fi
\providecommand{\url}[1]{\href{#1}{#1}}
\providecommand{\dodoi}[1]{doi:~\href{http://doi.org/#1}{\nolinkurl{#1}}}
\providecommand{\doeprint}[1]{\href{http://ascl.net/#1}{\nolinkurl{http://ascl.net/#1}}}
\providecommand{\doarXiv}[1]{\href{https://arxiv.org/abs/#1}{\nolinkurl{https://arxiv.org/abs/#1}}}

\bibitem[{{Al-Haddad} {et~al.}(2011){Al-Haddad}, {Roussev}, {M{\"o}stl},
  {Jacobs}, {Lugaz}, {Poedts}, \& {Farrugia}}]{AlHaddad:2011}
{Al-Haddad}, N., {Roussev}, I.~I., {M{\"o}stl}, C., {et~al.} 2011, The
  Astrophys. Journ. Lett., 738, L18, \dodoi{10.1088/2041-8205/738/2/L18}

\bibitem[{{Ala-Lahti} {et~al.}(2020){Ala-Lahti}, {Ruohotie}, {Good}, {Kilpua},
  \& {Lugaz}}]{AlaLahti:2020}
{Ala-Lahti}, M., {Ruohotie}, J., {Good}, S., {Kilpua}, E.~K.~J., \& {Lugaz}, N.
  2020, Journal of Geophysical Research: Space Physics, 125,
  \dodoi{10.1029/2020JA028002}

\bibitem[{{Chree}(1913)}]{Chree:1913}
{Chree}, C. 1913, Philosophical Transactions of the Royal Society of London.
  Series A, Containing Papers of a Mathematical or Physical Character, 212, 75,
  \dodoi{10.1098/rsta.1913.0003}

\bibitem[{{Das} {et~al.}(2011){Das}, {Opher}, {Evans}, {Loesch}, \&
  {Gombosi}}]{Das:2011}
{Das}, I., {Opher}, M., {Evans}, R., {Loesch}, C., \& {Gombosi}, T.~I. 2011,
  The Astrophysical Journal, 729, \dodoi{10.1088/0004-637X/729/2/112}

\bibitem[{{D{\'e}moulin} \& {Dasso}(2009)}]{Demoulin:2009}
{D{\'e}moulin}, P., \& {Dasso}, S. 2009, Astron. Astrophys., 498, 551

\bibitem[{{D{\'e}moulin} {et~al.}(2020){D{\'e}moulin}, {Dasso}, {Lanabere}, \&
  {Janvier}}]{Demoulin:2020}
{D{\'e}moulin}, P., {Dasso}, S., {Lanabere}, V., \& {Janvier}, M. 2020, Astron.
  Astrophys., 639, A6, \dodoi{10.1051/0004-6361/202038077}

\bibitem[{{D{\'e}moulin} {et~al.}(2008){D{\'e}moulin}, {Nakwacki}, {Dasso}, \&
  {Mandrini}}]{Demoulin:2008}
{D{\'e}moulin}, P., {Nakwacki}, M.~S., {Dasso}, S., \& {Mandrini}, C.~H. 2008,
  Solar Phys., 250, 347, \dodoi{10.1007/s11207-008-9221-9}

\bibitem[{{Echer} {et~al.}(2008){Echer}, {Gonzalez}, {Tsurutani}, \&
  {Gonzalez}}]{Echer:2008}
{Echer}, E., {Gonzalez}, W.~D., {Tsurutani}, B.~T., \& {Gonzalez}, A.~L.~C.
  2008, J. Geophys. Res., 113, A05221, \dodoi{10.1029/2007JA012744}

\bibitem[{{Farrugia} {et~al.}(1993){Farrugia}, {Freeman}, {Burlaga}, {Lepping},
  \& {Takahashi}}]{Farrugia:1993b}
{Farrugia}, C.~J., {Freeman}, M.~P., {Burlaga}, L.~F., {Lepping}, R.~P., \&
  {Takahashi}, K. 1993, J. Geophys. Res., 98, 7657, \dodoi{10.1029/92JA02351}

\bibitem[{{Feng} \& {Wang}(2013)}]{Feng:2013}
{Feng}, H., \& {Wang}, J. 2013, Astron. Astrophys., 559, A92,
  \dodoi{10.1051/0004-6361/201322522}

\bibitem[{{Fox} {et~al.}(2016){Fox}, {Velli}, \& {Bale}}]{Fox:2016}
{Fox}, N.~J., {Velli}, M.~C., \& {Bale}, S.~D. 2016, Space Science Reviews,
  204, 7, \dodoi{10.1007/s11214-015-0211-6}

\bibitem[{{Galvin} {et~al.}(2008){Galvin}, {Kistler}, {Popecki}, {Farrugia},
  {Simunac}, {Ellis}, {M{\"o}bius}, {Lee}, {Boehm}, {Carroll}, {Crawshaw},
  {Conti}, {Demaine}, {Ellis}, {Gaidos}, {Googins}, {Granoff}, {Gustafson},
  {Heirtzler}, {King}, {Knauss}, {Levasseur}, {Longworth}, {Singer}, {Turco},
  {Vachon}, {Vosbury}, {Widholm}, {Blush}, {Karrer}, {Bochsler}, {Daoudi},
  {Etter}, {Fischer}, {Jost}, {Opitz}, {Sigrist}, {Wurz}, {Klecker}, {Ertl},
  {Seidenschwang}, {Wimmer-Schweingruber}, {Koeten}, {Thompson}, \&
  {Steinfeld}}]{Galvin:2008}
{Galvin}, A.~B., {Kistler}, L.~M., {Popecki}, M.~A., {et~al.} 2008, Space
  Science Reviews, 136, 437, \dodoi{10.1007/s11214-007-9296-x}

\bibitem[{{Good} {et~al.}(2020){Good}, {Ala-Lahti}, {Palmerio}, {Kilpua}, \&
  {Osmane}}]{Good:2020}
{Good}, S.~W., {Ala-Lahti}, M., {Palmerio}, E., {Kilpua}, E.~K.~J., \&
  {Osmane}, A. 2020, The Astrophysical Journal, 893,
  \dodoi{10.3847/1538-4357/ab7fa2}

\bibitem[{{Gopalswamy}(2006)}]{Gopalswamy:2006}
{Gopalswamy}, N. 2006, Space Science Reviews, 124, 145,
  \dodoi{10.1007/s11214-006-9102-1}

\bibitem[{{Green} {et~al.}(2018){Green}, {T{\"o}r{\"o}k}, \& {Vr{\v
  s}nak}}]{Green:2018}
{Green}, L.~M., {T{\"o}r{\"o}k}, T., \& {Vr{\v s}nak}, B. 2018, Space Science
  Reviews, 214, 52, \dodoi{10.1007/s11214-017-0462-5}

\bibitem[{{Gulisano} {et~al.}(2010){Gulisano}, {D{\'e}moulin}, {Dasso}, {Ruiz},
  \& {Marsch}}]{Gulisano:2010}
{Gulisano}, A.~M., {D{\'e}moulin}, P., {Dasso}, S., {Ruiz}, M.~E., \& {Marsch},
  E. 2010, Astron. Astrophys., 509, A39, \dodoi{10.1051/0004-6361/200912375}

\bibitem[{{Guo} {et~al.}(2010){Guo}, {Feng}, {Zhang}, {Zuo}, \&
  {Xiang}}]{Guo:2010}
{Guo}, J., {Feng}, X., {Zhang}, J., {Zuo}, P., \& {Xiang}, C. 2010, Journal of
  Geophysical Research: Space Physics, 115, \dodoi{10.1029/2009JA015140}

\bibitem[{{Hietala} {et~al.}(2014){Hietala}, {Kilpua}, {Turner}, \&
  {Angelopoulos}}]{Hietala:2014}
{Hietala}, H., {Kilpua}, E.~K.~J., {Turner}, D.~L., \& {Angelopoulos}, V. 2014,
  Geophys. Res. Lett., 41, 2258, \dodoi{10.1002/2014GL059551}

\bibitem[{{Janvier} {et~al.}(2019){Janvier}, {Winslow}, {Good}, {Bonhomme},
  {D{\'e}moulin}, \& {Dasso}}]{Janvier:2019}
{Janvier}, M., {Winslow}, R.~M., {Good}, S., {et~al.} 2019, Journal of
  Geophysical Research: Space Physics, 124, \dodoi{10.1029/2018JA025949}

\bibitem[{{Jian} {et~al.}(2006){Jian}, {Russell}, {Luhmann}, \&
  {Skoug}}]{Jian:2006}
{Jian}, L., {Russell}, C.~T., {Luhmann}, J.~G., \& {Skoug}, R.~M. 2006, Solar
  Phys., 239, 393, \dodoi{10.1007/s11207-006-0133-2}

\bibitem[{{Jian} {et~al.}(2011){Jian}, {Russell}, \& {Luhmann}}]{Jian:2011}
{Jian}, L.~K., {Russell}, C.~T., \& {Luhmann}, J.~G. 2011, Solar Phys., 274,
  321, \dodoi{10.1007/s11207-011-9737-2}

\bibitem[{{Jian} {et~al.}(2018){Jian}, {Russell}, {Luhmann}, \&
  {Galvin}}]{Jian:2018}
{Jian}, L.~K., {Russell}, C.~T., {Luhmann}, J.~G., \& {Galvin}, A.~B. 2018, The
  Astrophysical Journal, 885, 114, \dodoi{10.3847/1538-4357/aab189}

\bibitem[{{Kaiser}(2005)}]{Kaiser:2005}
{Kaiser}, M.~L. 2005, Advances in Space Research, 36, 1483

\bibitem[{{Kataoka} {et~al.}(2005){Kataoka}, {Watari}, {Shimada}, {Shimazu}, \&
  {Marubashi}}]{Kataoka:2005}
{Kataoka}, R., {Watari}, S., {Shimada}, N., {Shimazu}, H., \& {Marubashi}, K.
  2005, Geophys. Res. Lett., 32, L12103, \dodoi{10.1029/2005GL022777}

\bibitem[{{Kaymaz} \& {Siscoe}(2006)}]{Kaymaz:2006}
{Kaymaz}, Z., \& {Siscoe}, G. 2006, Solar Phys., 239, 437,
  \dodoi{10.1007/s11207-006-0308-x}

\bibitem[{{Kilpua} {et~al.}(2019){Kilpua}, {Fontaine}, {Moissard},
  {Ala-Lahti,}, {Palmerio}, \& {Yordanova}}]{Kilpua:2019b}
{Kilpua}, E.~K.~J., {Fontaine}, D., {Moissard}, C., {et~al.} 2019, Space
  Weather, 17, 1257, \dodoi{10.1029/2019SW002217}

\bibitem[{{Kilpua} {et~al.}(2013){Kilpua}, {Hietala}, {Koskinen}, {Fontaine},
  \& {Turc}}]{Kilpua:2013}
{Kilpua}, E.~K.~J., {Hietala}, H., {Koskinen}, H.~E.~J., {Fontaine}, D., \&
  {Turc}, L. 2013, Annales Geophysicae, 31, 1559,
  \dodoi{10.5194/angeo-31-1559-2013}

\bibitem[{{Kilpua} {et~al.}(2011){Kilpua}, {Jian}, {Li}, {Luhmann}, \&
  {Russell}}]{Kilpua:2011}
{Kilpua}, E.~K.~J., {Jian}, L.~K., {Li}, Y., {Luhmann}, J.~G., \& {Russell},
  C.~T. 2011, J. Atmos. Sol. Terr. Phys., 73, 1228,
  \dodoi{10.1016/j.jastp.2010.10.012}

\bibitem[{{Kilpua} {et~al.}(2017){Kilpua}, {Koskinen}, \&
  {Pulkkinen}}]{Kilpua:2017b}
{Kilpua}, E.~K.~J., {Koskinen}, H.~E.~J., \& {Pulkkinen}, T.~I. 2017, Living
  Reviews in Solar Physics, 14, \dodoi{10.1007/s41116-017-0009-6}

\bibitem[{{Kilpua} {et~al.}(2015){Kilpua}, {Lumme}, {Andreeova}, {Isavnin}, \&
  {Koskinen}}]{Kilpua:2015a}
{Kilpua}, E.~K.~J., {Lumme}, E., {Andreeova}, K., {Isavnin}, A., \& {Koskinen},
  H.~E.~J. 2015, Journal of Geophysical Research: Space Physics, 120, 4112,
  \dodoi{10.1002/2015JA021138}

\bibitem[{{Kilpua} {et~al.}(2020){Kilpua}, {Fontaine}, {Good}, {Ala-Lahti},
  {Osmane}, {Palmerio}, {Yordanova}, {Moissard}, {Hadid}, \&
  {Janvier}}]{Kilpua:2020}
{Kilpua}, E.~K.~J., {Fontaine}, D., {Good}, S.~W., {et~al.} 2020, Annales
  Geophysicae, 38, 999, \dodoi{10.5194/angeo-38-999-2020}

\bibitem[{{Liu} {et~al.}(2020){Liu}, {Chen}, \& {Zhao}}]{Liu:2020}
{Liu}, Y.~D., {Chen}, C., \& {Zhao}, X. 2020, The Astrophys. Journ. Lett., 897,
  \dodoi{10.3847/2041-8213/ab9d25}

\bibitem[{{Lugaz} {et~al.}(2018){Lugaz}, {Farrugia}, {Winslow}, {Al-Haddad},
  {Galvin}, {Nieves-Chinchilla}, {Lee}, \& {Janvier}}]{Lugaz:2018}
{Lugaz}, N., {Farrugia}, C.~J., {Winslow}, R.~M., {et~al.} 2018, The Astrophys.
  Journ. Lett., 864, 6, \dodoi{10.3847/2041-8213/aad9f4}

\bibitem[{{Lugaz} {et~al.}(2016){Lugaz}, {Farrugia}, {Winslow}, {Al-Haddad},
  {Kilpua}, \& {Riley}}]{Lugaz:2016}
---. 2016, Journal of Geophysical Research: Space Physics, 121,
  \dodoi{10.1002/2016JA023100}

\bibitem[{{Lugaz} {et~al.}(2017){Lugaz}, {Farrugia}, {Winslow}, {Small},
  {Manion}, \& {Savani}}]{Lugaz:2017a}
---. 2017, The Astrophysical Journal, 848, 75, \dodoi{10.3847/1538-4357/aa8ef9}

\bibitem[{{Lugaz} {et~al.}(2020{\natexlab{a}}){Lugaz}, {Salman}, {Winslow},
  {Al-Haddad}, {Farrugia}, {Zhuang}, \& {Galvin}}]{Lugaz:2020b}
{Lugaz}, N., {Salman}, T.~M., {Winslow}, R.~M., {et~al.} 2020{\natexlab{a}},
  The Astrophysical Journal, 899, \dodoi{10.3847/1538-4357/aba26b}

\bibitem[{{Lugaz} {et~al.}(2020{\natexlab{b}}){Lugaz}, {Winslow}, \&
  {Farrugia}}]{Lugaz:2020a}
{Lugaz}, N., {Winslow}, R.~M., \& {Farrugia}, C.~J. 2020{\natexlab{b}}, Journal
  of Geophysical Research: Space Physics, 125, \dodoi{10.1029/2019JA027213}

\bibitem[{{Luhmann} {et~al.}(2008){Luhmann}, {Curtis}, {Schroeder}, {McCauley},
  {Lin}, {Larson}, {Bale}, {Sauvaud}, {Aoustin}, {Mewaldt}, {Cummings},
  {Stone}, {Davis}, {Cook}, {Kecman}, {Wiedenbeck}, {von Rosenvinge}, {Acuna},
  {Reichenthal}, {Shuman}, {Wortman}, {Reames}, {Mueller-Mellin}, {Kunow},
  {Mason}, {Walpole}, {Korth}, {Sanderson}, {Russell}, \&
  {Gosling}}]{Luhmann:2008}
{Luhmann}, J.~G., {Curtis}, D.~W., {Schroeder}, P., {et~al.} 2008, Space
  Science Reviews, 136, 117, \dodoi{10.1007/s11214-007-9170-x}

\bibitem[{{Manchester} {et~al.}(2017){Manchester}, {Kilpua}, {Liu}, {Lugaz},
  {Riley}, {T{\"o}r{\"o}k}, \& {Vr{\v s}nak}}]{Manchester:2017}
{Manchester}, W., {Kilpua}, E.~K.~J., {Liu}, Y.~D., {et~al.} 2017, Space
  Science Reviews, 212, 1159, \dodoi{10.1007/s11214-017-0394-0}

\bibitem[{{Manchester} {et~al.}(2005){Manchester}, {Gombosi}, {De Zeeuw},
  {Sokolov}, {Roussev}, {Powell}, {K{\' o}ta}, {T{\' o}th}, \&
  {Zurbuchen}}]{Manchester:2005}
{Manchester}, W.~B., {Gombosi}, T.~I., {De Zeeuw}, D.~L., {et~al.} 2005, The
  Astrophysical Journal, 622, 1225

\bibitem[{{Mas{\'i}as-Meza} {et~al.}(2016){Mas{\'i}as-Meza}, {Dasso},
  {D{\'e}moulin}, {Rodriguez}, \& {Janvier}}]{Masias:2016}
{Mas{\'i}as-Meza}, J.~J., {Dasso}, S., {D{\'e}moulin}, P., {Rodriguez}, L., \&
  {Janvier}, M. 2016, Astron. Astrophys., 592,
  \dodoi{10.1051/0004-6361/201628571}

\bibitem[{{Mitsakou} \& {Moussas}(2014)}]{Mitsakou:2014}
{Mitsakou}, E., \& {Moussas}, X. 2014, Solar Phys., 289, 3137,
  \dodoi{10.1007/s11207-014-0505-y}

\bibitem[{{Moissard} {et~al.}(2019){Moissard}, {Fontaine}, \&
  {Savoini}}]{Moissard:2019}
{Moissard}, C., {Fontaine}, D., \& {Savoini}, P. 2019, Journal of Geophysical
  Research: Space Physics, 124, 8208, \dodoi{10.1029/2019JA026952}

\bibitem[{{M{\"u}ller} {et~al.}(2013){M{\"u}ller}, {Marsden}, {St.~Cyr}, \&
  {Gilbert}}]{Mueller:2013}
{M{\"u}ller}, D., {Marsden}, R.~G., {St.~Cyr}, O.~C., \& {Gilbert}, H.~R. 2013,
  Solar Phys., 285, 25, \dodoi{10.1007/s11207-012-0085-7}

\bibitem[{{Nieves-Chinchilla} {et~al.}(2018){Nieves-Chinchilla}, {Vourlidas},
  \& {Raymond}}]{Nieves:2018}
{Nieves-Chinchilla}, T., {Vourlidas}, A., \& {Raymond}, J.~C. 2018, Solar
  Phys., 293, \dodoi{10.1007/s11207-018-1247-z}

\bibitem[{{Owens}(2020)}]{Owens:2020}
{Owens}, M.~J. 2020, Solar Phys., 295, 148, \dodoi{10.1007/s11207-020-01721-0}

\bibitem[{{Owens} {et~al.}(2005){Owens}, {Cargill}, {Pagel}, {Siscoe}, \&
  {Crooker }}]{Owens:2005}
{Owens}, M.~J., {Cargill}, P.~J., {Pagel}, C., {Siscoe}, G.~L., \& {Crooker },
  N.~U. 2005, J. Geophys. Res., 110, A01105, \dodoi{10.1029/2004JA010814}

\bibitem[{{Owens} {et~al.}(2017){Owens}, {Lockwood}, \&
  {Barnard}}]{Owens:2017a}
{Owens}, M.~J., {Lockwood}, M., \& {Barnard}, L.~A. 2017, Scientific Reports,
  7, 4152, \dodoi{10.1038/s41598-017-04546-3}

\bibitem[{{Paulson} {et~al.}(2012){Paulson}, {Taylor}, {Smith}, {Vasquez}, \&
  {Hu}}]{Paulson:2012}
{Paulson}, K.~W., {Taylor}, D.~K., {Smith}, C.~W., {Vasquez}, B.~J., \& {Hu},
  Q. 2012, Space Weather, 10, \dodoi{10.1029/2012SW000855}

\bibitem[{{Regnault} {et~al.}(2020){Regnault}, {Janvier}, {D{\'e}moulin},
  {Auch{\`e}re}, {Strugarek}, {Dasso}, \& {No{\^u}s}}]{Regnault:2020}
{Regnault}, F., {Janvier}, M., {D{\'e}moulin}, P., {et~al.} 2020, Journal of
  Geophysical Research: Space Physics, 125, \dodoi{10.1029/2020JA028150}

\bibitem[{{Richardson} \& {Cane}(2010)}]{Richardson:2010}
{Richardson}, I.~G., \& {Cane}, H.~V. 2010, Solar Phys., 264, 189,
  \dodoi{10.1007/s11207-010-9568-6}

\bibitem[{{Richardson} \& {Cane}(2012)}]{Richardson:2012}
---. 2012, J. Space Weather Space Clim, 2, \dodoi{10.1051/swsc/2012001}

\bibitem[{{Riley} \& {Richardson}(2012)}]{Riley:2012}
{Riley}, P., \& {Richardson}, I.~G. 2012, Solar Phys., 284, 217,
  \dodoi{10.1007/s11207-012-0006-9}

\bibitem[{{Rodriguez} {et~al.}(2016){Rodriguez}, {Mas{\'i}as-Meza}, \&
  {Dasso}}]{Rodriguez:2016}
{Rodriguez}, L., {Mas{\'i}as-Meza}, J.~J., \& {Dasso}, S. 2016, Solar Phys.,
  291, 2145, \dodoi{10.1007/s11207-016-0955-5}

\bibitem[{{Ruffenach} {et~al.}(2015){Ruffenach}, {Lavraud}, {Farrugia},
  {D{\'e}moulin}, {Dasso}, {Owens}, {Sauvaud}, {Rouillard}, {Lynnyk},
  {Foullon}, {Savani}, {Luhmann}, \& {Galvin}}]{Ruffenach:2015}
{Ruffenach}, A., {Lavraud}, B., {Farrugia}, C.~J., {et~al.} 2015, J. Geophys.
  Res., 120, 43, \dodoi{10.1002/2014JA020628}

\bibitem[{{Russell} \& {Mulligan}(2002)}]{Russell:2002}
{Russell}, C.~T., \& {Mulligan}, T. 2002, Planetary and Space Science, 50, 527,
  \dodoi{10.1016/s0032-0633(02)00031-4}

\bibitem[{{Salman} {et~al.}(2018){Salman}, {Lugaz}, {Farrugia}, {Winslow},
  {Galvin}, \& {Schwadron}}]{Salman:2018}
{Salman}, T.~M., {Lugaz}, N., {Farrugia}, C.~J., {et~al.} 2018, Space Weather,
  16, \dodoi{10.1029/2018SW002056}

\bibitem[{{Salman} {et~al.}(2020{\natexlab{a}}){Salman}, {Lugaz}, {Farrugia},
  {Winslow}, {Jian}, \& {Galvin}}]{Salman:2020b}
---. 2020{\natexlab{a}}, The Astrophysical Journal, 904,
  \dodoi{10.3847/1538-4357/abbdf5}

\bibitem[{{Salman} {et~al.}(2020{\natexlab{b}}){Salman}, {Winslow}, \&
  {Lugaz}}]{Salman:2020a}
{Salman}, T.~M., {Winslow}, R.~M., \& {Lugaz}, N. 2020{\natexlab{b}}, Journal
  of Geophysical Research: Space Physics, 125, \dodoi{10.1029/2019JA027084}

\bibitem[{{Siscoe} \& {Odstrcil}(2008)}]{Siscoe:2008}
{Siscoe}, G., \& {Odstrcil}, D. 2008, J. Geophys. Res., 113, A00B07,
  \dodoi{10.1029/2008JA013142}

\bibitem[{{Temmer} {et~al.}(2021){Temmer}, {Dumbovi{\'c}}, {Holzknecht}, {Vr{\v
  s}nak}, {Sachdeva}, \& {Heinemann}}]{Temmer:2021}
{Temmer}, M., {Dumbovi{\'c}}, M., {Holzknecht}, L., {et~al.} 2021, Journal of
  Geophysical Research: Space Physics, 126, \dodoi{10.1029/2020JA028380}

\bibitem[{{Turner} {et~al.}(2019){Turner}, {Kilpua}, {Hietala}, {Claudepierre},
  {O'Brien}, \& {Fennell}}]{Turner:2019}
{Turner}, D.~L., {Kilpua}, E.~K.~J., {Hietala}, H., {et~al.} 2019, Journal of
  Geophysical Research: Space Physics, 124, 1013, \dodoi{10.1029/2018JA026066}

\bibitem[{{Vourlidas} {et~al.}(2018){Vourlidas}, {Fischer}, \&
  {Webb}}]{Vourlidas:2018}
{Vourlidas}, A., {Fischer}, S., \& {Webb}, D.~F. 2018, The Astrophysical
  Journal, 861, 11, \dodoi{10.3847/1538-4357/aaca3e}

\bibitem[{{Vr{\v s}nak} {et~al.}(2013){Vr{\v s}nak}, {{\v Z}ic}, {Vrbanec},
  {Temmer}, {Rollett}, {M{\"o}stl}, {Veronig}, {{\v C}alogovi{\'c}},
  {Dumbovi{\'c}}, {Luli{\'c}}, {Moon}, \& {Shanmugaraju}}]{Vrsnak:2013}
{Vr{\v s}nak}, B., {{\v Z}ic}, T., {Vrbanec}, D., {et~al.} 2013, Solar Phys.,
  285, 295, \dodoi{10.1007/s11207-012-0035-4}

\bibitem[{{Vr{\v s}nak} {et~al.}(2019){Vr{\v s}nak}, {Amerstorfer},
  {Dumbovi{\'c}}, {Leitner}, {Veronig}, {Temmer}, {M{\"o}stl}, {Amerstorfer},
  {Farrugia}, \& {Galvin}}]{Vrsnak:2019}
{Vr{\v s}nak}, B., {Amerstorfer}, T., {Dumbovi{\'c}}, M., {et~al.} 2019, The
  Astrophysical Journal, 877, \dodoi{10.3847/1538-4357/ab190a}

\bibitem[{{Winslow} {et~al.}(2015){Winslow}, {Lugaz}, {Philpott}, {Schwadron},
  {Farrugia}, {Anderson}, \& {Smith}}]{Winslow:2015}
{Winslow}, R.~M., {Lugaz}, N., {Philpott}, L.~C., {et~al.} 2015, Journal of
  Geophysical Research: Space Physics, 120, 6101, \dodoi{10.1002/2015JA021200}

\bibitem[{{Zhang} {et~al.}(2007){Zhang}, {Richardson}, {Webb}, {Gopalswamy},
  {Huttunen}, {Kasper}, {Nitta}, {Poomvises}, {Thompson}, {Wu}, {Yashiro}, \&
  {Zhukov}}]{Zhang:2007}
{Zhang}, J., {Richardson}, I.~G., {Webb}, D.~F., {et~al.} 2007, J. Geophys.
  Res., 112, A10102, \dodoi{10.1029/2007JA012321}

\end{thebibliography}
\bibliographystyle{aasjournal} 

\end{document}